 \documentclass{aa}
\usepackage{graphicx}
\usepackage{color}
\usepackage{txfonts}
\usepackage{natbib}
\bibpunct{(}{)}{;}{a}{}{,} 

\begin{document}
\newcommand{\kep}{{{\it Kepler }}}
\title{Magnetic activity and differential rotation in the young Sun-like stars KIC\,7985370 and 
KIC\,7765135\thanks{Based on public \kep data, on 
observations made with the Italian Telescopio Nazionale Galileo (TNG) operated by the Fundaci\'on 
Galileo Galilei -- INAF at the Observatorio del Roque del los Muchachos, La Palma (Canary Islands), 
on observations collected at the 2.2-m telescope of the Centro Astron\'omico Hispano Alem\'an (CAHA) 
at Calar Alto (Almer\'{\i}a, Spain), operated jointly by the Max-Planck-Institut f\"ur Astronomie and 
the Instituto de Astrof\'{\i}sica de Andaluc\'{\i}a (CSIC),
and on observations collected at the Catania Astrophysical Observatory (Italy).}}

\author{H.-E. Fr\"ohlich\inst{1}\and
	A. Frasca\inst{2}\and
	G. Catanzaro\inst{2}\and
	A. Bonanno \inst{2}\and
	E. Corsaro\inst{2,3}\and
	J. Molenda-\.Zakowicz\inst{4}\and
	A. Klutsch\inst{5}\and
	D. Montes\inst{5}
	        }

\offprints{H.-E. Fr\"ohlich\\ \email{hefroehlich@aip.de}}

\institute{
Leibniz Institute for Astrophysics Potsdam (AIP), An der Sternwarte 16, 14482 Potsdam, Germany
\and
INAF, Osservatorio Astrofisico di Catania, via S. Sofia, 78, 95123 Catania, Italy
\and
Universit\`{a} di Catania, Dipartimento di Fisica e Astronomia,  via S. Sofia, 78, 95123 Catania, Italy
\and 
Astronomical Institute, Wroc{\l}aw University, ul.\ Kopernika 11, 51-622 Wroc{\l}aw, Poland
\and
Departamento de Astrof\'{\i}sica y Ciencias de la Atm\'osfera, Universidad Complutense de Madrid, 28040 Madrid, Spain
}

\date{Received  / accepted}

 
\abstract 
{}
{We present a detailed study of the two Sun-like stars \object{KIC\,7985370} and \object{KIC\,7765135}, 
aimed at determining their activity level, spot distribution, and differential rotation. Both stars were discovered by 
us to be young stars and were observed by the NASA \kep mission.}
{The fundamental stellar parameters ($v \sin i$, spectral type, $T_{\rm eff}$, $\log g$, and [Fe/H]) were derived from 
optical spectroscopy by the comparison with both standard-star and synthetic spectra. The spectra of the targets allowed us also to 
study the chromospheric activity from the emission in the core of hydrogen H$\alpha$ and \ion{Ca}{ii} infrared triplet (IRT) lines, 
revealed by the subtraction of inactive templates. 
The high-precision Kepler photometric data spanning over 229 days were then fitted with a robust spot model. 
Model selection and parameter estimation are performed in a 
Bayesian manner, using a Markov chain Monte Carlo method.}
{Both stars came out to be Sun-like (G1.5\,V spectral type) with an age of about 100--200 Myr,
based on their lithium content and kinematics. Their youth is confirmed by the high level of chromospheric 
activity, which is comparable to that displayed by the early G-type stars in the Pleiades cluster. 
The Balmer decrement and flux ratio of the \ion{Ca}{ii}-IRT lines suggest
that the formation of the core of these lines occurs mainly in optically-thick regions that are analogous to solar plages. 
The spot model applied to the \kep photometry requires at least seven enduring spots in the case of KIC\,7985370 and nine spots 
in the case of KIC\,7765135 for a satisfactory fit of the data. 
The assumption of longevity of the star spots, whose area is allowed to evolve in time, is at the heart of our spot-modelling approach. 
On both stars the surface differential rotation is Sun-like, with the high-latitude 
spots rotating slower than the low-latitude ones. {We found, for both stars, a rather high value of the equator-to-pole differential 
rotation (${\rm d}\Omega\approx 0.18$ rad\,d$^{-1}$) which is in contrast with the predictions of some mean-field models of differential 
rotation for fast-rotating stars. Our results are instead in agreement with previous works on solar-type stars and 
with other models which predict a higher latitudinal shear, increasing with equatorial angular velocity, that can 
undergo changes along the magnetic cycle.} } 
{}

\titlerunning{Magnetic activity and differential rotation in two young suns}
      \authorrunning{H.-E. Fr\"ohlich et al.}

\maketitle

\section{Introduction}
\label{Sec:intro}

In the Sun, magnetic activity is thought to be produced by a global-scale dynamo 
action arising from the coupling 
of convection and rotation \citep{Parker1955,Steenbecketal1966}. Young Sun-like stars are rotating faster than 
the Sun and display a much higher level of magnetic activity at all atmospheric
layers, which is likely due to a stronger dynamo action.
They also show different manifestations of activity compared to the Sun, 
such as bigger and long-living spots in their photospheres, active longitude belts,
absence or a different behaviour of activity cycles, highly energetic flares, etc.
These differences are likely related to the dynamo mechanism, which is operating in
rather different conditions in young stars mainly regarding their rotation rate and 
internal structure. Understanding the properties of young suns 
and, particularly, their activity and rotation, is crucial to trace the Sun and
its environment back to the first evolutionary stages.


In fact the properties of the magnetoconvection in these stars seems to be
strongly influenced by the $\Omega$-effect that produces characteristic
``wreaths'' on large scales \citep{Nelson11}. Although strong
latitudinal variations
of the differential rotation can be obtained by means of the combined
role of thermal  wind balance and geostrophy, the results of numerical
simulations seem to be strongly dependent on the Reynolds number of
the flow. The situation is at variance with the dynamo action in main
sequence solar-type stars, where the role of the tachocline is instead
essential in producing the $\alpha$ effect \citep{Dikpati2001,Bonanno02,Bonanno2012}.

It is still unclear whether a strong latitudinal differential rotation is common among fast-rotating stars.
\citet{Marsden11} report values of the absolute differential rotation ${\rm d}\Omega$ in the range 0.08--0.45 rad\,d$^{-1}$
for a sample of rapid rotators similar to and slightly more massive than the Sun. Despite the spread of values, it seems that ${\rm d}\Omega$ is in any case larger than in
the Sun. 
The measures of absolute differential rotation in a large sample of F- and early G-type stars
through the Fourier transform technique \citep{ReinersSchmitt2003,Reiners2006} show no indication of the decrease in 
this parameter with the rotation period, rather the highest values of ${\rm d}\Omega$ are encountered for periods 
between 2 and 3 days. 
On the other hand, some recent calculations predict a moderate differential rotation, comparable to that of the Sun, also
for a Sun-like star rotating 20 times faster \citep[e.g.,][]{Kueker11}. 

Moreover, ${\rm d}\Omega$ seems also to be a function of the stellar mass for main-sequence stars, increasing with their effective temperature, as shown, e.g., by 
\citet{Barnes05}. One of the largest values of differential rotation for a star noticeably cooler than the Sun 
was found by us \citep[][hereafter Paper~I]{Frasca2011} for \object{KIC~8429280}, a 50\,Myr-old K2-type star, from the analysis of the light curve collected by the NASA \kep spacecraft.

The highly precise photometry of \kep \citep{Borucki10,Koch10} coupled with the long and virtually uninterrupted coverage makes these data unique for
the study of photospheric activity and differential rotation in late-type stars, as we have shown in 
our first work based on the analysis of \kep data of spotted stars (Paper~I). 

However, whether star spots are indeed the best tracers of the surface rotation or not is still a matter of debate  
\cite[for a different point of view, see][]{KorhonenElstner2011}. 


  As for \object{KIC\,8429280} (Paper~I), the two new targets 
  \object{KIC\,7985370} (\object{HD\,189210}~=~\object{2MASS\,J19565974+4345083}~= \object{TYC\,3149-1571-1}) 
  and \object{KIC\,7765135} (\object{2MASS\,J19425057 +4324486} = \object{TYC\,3148-2163-1}) 
were selected as active stars from their optical variability and from the cross-correlation of the ROSAT All-Sky Survey 
(RASS; \citealt{Voges1999, Voges2000}) with Tycho and Hipparcos catalogues \citep{HIPPA97}. 
{With $V=10\fm0$ and $11\fm8$, respectively, both stars are relatively bright ones in the \kep field of view. }
Both of them were recently reported as variable by \citet{Pigulski2009}, who searched for bright variable stars in the \kep 
field of view with ASAS3-North station. The variability of KIC\,7985370 could be due to a rotational modulation according to 
\citet{Uytte2011} that rely on the first two quarters of \kep data. 
The \kep light curves readily show these stars as rotationally variable with a period of about 2--3 days, which is typical of G-type stars in the Pleiades
cluster (age $\approx$\,130\,Myr, \citealt{Barrado2004}). 
The estimates of their atmospheric parameters reported in the \kep Input Catalog (KIC), which are based on Sloan photometry 
\citep[for a revised temperature scale cf.][]{Pinsonneault2012}, 
suggested to us that these objects were similar to the Sun. 

The analysis of the optical spectra collected by us confirmed that the stars are nearly identical to the Sun, but much younger and 
as such deserving a detailed investigation.
Applying the same techniques as in Paper~I, we determined the basic stellar parameters 
(Sect.\,\ref{Sec:Analysis}), the chromospheric activity (Sect.\,\ref{Sec:Chrom}), and elements abundances (Sect.\,\ref{Sec:Abundance}). 
The kinematics of these stars is shortly discussed in Sect.\,\ref{Sec:Kinematics}. 
The Bayesian approach to spot modelling applied to the \kep light curves is described in Sect.\,\ref{Sec:Spot_mod} and the results are presented and discussed 
in Sect.\,\ref{Sec:Disc} with particular relevance to the issue of differential rotation.

\section{Ground-based observations and data reduction}
\label{Sec:Data}

\subsection{Spectroscopy}
Two spectra of KIC\,7985370 with a signal-to-noise ratio S/N of about 60--70 were collected at the {\em M. G. Fracastoro\/} station (Serra  
La Nave, Mt. Etna, 1750 m a.s.l.) of the \textit{Osservatorio Astrofisico di Catania} (OAC, Italy) in July and October 2009. 
The 91-cm telescope of the OAC was equipped with FRESCO, a fiber-fed {\it \'echelle} spectrograph that, with the 300-lines/mm 
cross-disperser, covers the spectral range 4300--6800\,\AA\  with a resolution $R=\lambda/\Delta\lambda\approx\,21\,000$. 


Another spectrum of KIC\,7985370 with S/N$\approx$\,120 was taken with SARG, the {\it \'echelle} spectrograph at the Italian {\it Telescopio
Nazionale Galileo} (TNG, La Palma, Spain) on 2009 August 12 with the red grism and the slit width of 0$\farcs$8. This spectrum, covering the
5500--11\,000\,\AA\  wavelength range, has a resolution $R\approx\,57\,000$.

The spectrum of KIC\,7765135 was taken on 2009 October 3 with the Fiber Optics Cassegrain {\it \'Echelle} Spectrograph 
(FOCES; \citealt{Pfeiffer1998}) at the 2.2-m telescope of the Calar Alto Astronomical Observatory (CAHA, Almer\'{\i}a, 
Spain). The 2048$\times$2048 CCD detector Site\#1d (pixel size = 24\,$\mu$m) and the slit width of 400\,$\mu$m give rise to a 
resolution $R\approx\,28\,000$ and a S/N$\approx$\,100 in the red wavelength region with a 30-minutes exposure time.

{Spectra of radial (RV) and rotational velocity ($v\sin i$) standard stars (Table\,\ref{Tab:Standards}), as well as bias, 
flat-field, and arc-lamp exposures were acquired during each observing run and were used for the data reduction and analysis.} 

The data reduction was performed with the \textsc{Echelle} task of the IRAF\footnote{IRAF is distributed by the 
National Optical Astronomy Observatory, which is operated by the Association of the Universities for Research in 
Astronomy, inc. (AURA) under cooperative agreement with the National Science Foundation.} package
\citep[see, e.g.,][for details]{Frasca2010,Catanza10}.

\begin{table}[htb]
\caption{ Radial/rotational velocity standard stars.}
\label{Tab:Standards}
\centering
\begin{tabular}{llccl}
\hline
\hline
Name       & Sp. Type &   $RV$             & $v\sin i$\tablefootmark{c}      & Notes\tablefootmark{d}\\  
           &          &   \multicolumn{1}{c}{(km\,s$^{-1}$)}   & \multicolumn{1}{c}{(km\,s$^{-1}$)} &      \\  
\hline
\noalign{\smallskip}
\object{HD\,10307}  &  G1.5\,V & $4.0$\tablefootmark{b}\tablefootmark{\mathrm{*}} & 2.1  & $v\sin i$ \\  
\object{HD\,10700}  &  G8\,V   & $-17.1$\tablefootmark{b} & 0.9  & $v\sin i$ \\  
\object{HD\,157214} &  G0\,V   & $-79.2$\tablefootmark{b} & 1.6  & $v\sin i$ \\  
\object{HD\,182572} &  G8\,IV  & $-100.35$\tablefootmark{a} & 1.9  & $RV$, $v\sin i$ \\    
\object{HD\,187691} &  F8\,V   & $0.0$\tablefootmark{a}   & 2.8  & $RV$ \\  
\hline
\end{tabular}
\tablefoot{
\tablefoottext{a}{\citet{Udry}.}
\tablefoottext{b}{\citet{Nordstr}.}
\tablefoottext{c}{\citet{Glebocki2005}.}
\tablefoottext{d}{$RV=$ radial velocity standard star; $v\sin i=$ standard for rotational velocity.}
\tablefoottext{\mathrm{*}}{Variable?}
}
\end{table}

\subsection{Photometry}

The focal-reducer CCD camera at the 91-cm telescope of OAC was used, on 2009 December 10, to perform standard
photometry in the Johnson-Cousins $B$, $V$, $R_{\rm C}$, and $I_{\rm C}$ bands.  
The data were reduced following standard steps of overscan region subtraction, master-bias subtraction, 
and division by average twilight flat-field images. 
The $BVR_{\rm C}I_{\rm C}$ magnitudes were extracted from the corrected images through aperture photometry 
performed with DAOPHOT by using the IDL\footnote{IDL (Interactive Data Language) is a registered trademark 
of ITT Visual Information Solutions.} routine \textsc{APER}. 
Standard stars in the cluster \object{NGC~7790} and in the field of BL~Lac \citep{Stetson2000} were used to 
calculate the zero points and the transformation coefficients to the Johnson-Cousins system. 
Photometric data derived in this work and $JHK_{s}$ magnitudes from 2MASS \citep{2MASS} are summarized in Table~\ref{Tab:StarChar}.


\begin{table}[t]
\caption{Stellar parameters. The data of the upper part is from the literature.} 
\label{Tab:StarChar}
\centering
\begin{tabular}{llrclrcl}
\hline
\hline
\noalign{\smallskip}
                        && \multicolumn{3}{c}{KIC 7985370}&\multicolumn{3}{c}{KIC\,7765135}\\
\hline
\noalign{\smallskip}
RA                      &(J2000)      & \multicolumn{3}{c}{$19^{\rm h} 56^{\rm m} 59\fs 74$}  & \multicolumn{3}{c}{$19^{\rm h} 42^{\rm m} 50\fs 58$} \\
DEC                     &(J2000)      & \multicolumn{3}{c}{$+43\degr 45\arcmin 08\farcs 3$}   & \multicolumn{3}{c}{$+43\degr 24\arcmin 48\farcs 7$}  \\
X-ray count\tablefootmark{a} & [ct/s] & 0.033   &\hspace{- 8pt}$\pm$&\hspace{- 8pt}0.008      &   \multicolumn{3}{c}{\hspace{- 2pt}\dots{}}          \\                
$J$\tablefootmark{b}     &            &   8.789 &\hspace{- 8pt}$\pm$&\hspace{- 8pt}0.020      &  10.414 &\hspace{- 8pt}$\pm$&\hspace{- 8pt}0.025     \\
$H$\tablefootmark{b}     &            &   8.507 &\hspace{- 8pt}$\pm$&\hspace{- 8pt}0.063      &  10.113 &\hspace{- 8pt}$\pm$&\hspace{- 8pt}0.020     \\
$K_{\rm s}$\tablefootmark{b} &        &   8.419 &\hspace{- 8pt}$\pm$&\hspace{- 8pt}0.033      &  10.028 &\hspace{- 8pt}$\pm$&\hspace{- 8pt}0.014     \\
\noalign{\smallskip}
\hline
\noalign{\smallskip}
$B$                     &             &  10.59  &\hspace{- 8pt}$\pm$&\hspace{- 8pt}0.09       &  12.50  &\hspace{- 8pt}$\pm$&\hspace{- 8pt}0.09      \\                      
$V$                     &             &  9.98   &\hspace{- 8pt}$\pm$&\hspace{- 8pt}0.08       & 11.82   &\hspace{- 8pt}$\pm$&\hspace{- 8pt}0.08      \\
$R_{\rm C}$             &             &  9.57   &\hspace{- 8pt}$\pm$&\hspace{- 8pt}0.06       & 11.28   &\hspace{- 8pt}$\pm$&\hspace{- 8pt}0.06      \\
$I_{\rm C}$             &             &   9.33  &\hspace{- 8pt}$\pm$&\hspace{- 8pt}0.06       &  10.97  &\hspace{- 8pt}$\pm$&\hspace{- 8pt}0.06      \\
Sp. Type                &             & \multicolumn{3}{c}{\hspace{8pt}G1.5\,V}        & \multicolumn{3}{c}{\hspace{-2pt}G1.5\,V}         \\
$T_{\rm eff}$           &[K]          & 5815    &\hspace{- 8pt}$\pm$&\hspace{- 8pt}95         & 5835    &\hspace{- 8pt}$\pm$&\hspace{- 8pt}95        \\
$\log g$                &             & 4.24    &\hspace{- 8pt}$\pm$&\hspace{- 8pt}0.12       & 4.34    &\hspace{- 8pt}$\pm$&\hspace{- 8pt}0.12      \\
$[$Fe/H$]$              &             & $-0.05$ &\hspace{- 8pt}$\pm$&\hspace{- 8pt}0.10       & $0.04$  &\hspace{- 8pt}$\pm$&\hspace{- 8pt}0.05      \\
$v \sin i$\tablefootmark{c} &[km s$^{-1}$]& 18.2    &\hspace{- 8pt}$\pm$&\hspace{- 8pt}1.3        & 21.4    &\hspace{- 8pt}$\pm$&\hspace{- 8pt}1.1       \\
$v \sin i$\tablefootmark{d} &[km s$^{-1}$]& 17.9    &\hspace{- 8pt}$\pm$&\hspace{- 8pt}1.3        & 21.9    &\hspace{- 8pt}$\pm$&\hspace{- 8pt}1.3       \\
$RV$                    &[km s$^{-1}$]& $-24.0$ &\hspace{- 8pt}$\pm$&\hspace{- 8pt}0.3        & $-20.0$ &\hspace{- 8pt}$\pm$&\hspace{- 8pt}0.2       \\
$W$(Li)                 &[m\AA]       & 155     &\hspace{- 8pt}$\pm$&\hspace{- 8pt}20         & 160     &\hspace{- 8pt}$\pm$&\hspace{- 8pt}20        \\
$\log N$(Li)\tablefootmark{e}  &      & 2.87    &\hspace{- 8pt}$\pm$&\hspace{- 8pt}0.10       & 2.93    &\hspace{- 8pt}$\pm$&\hspace{- 8pt}0.10      \\
Age                     &[Myr]        & 100     &\hspace{-10pt} --  &\hspace{- 8pt}200        & 100     &\hspace{-10pt} --  &\hspace{- 8pt}200       \\       
Mass                    &[M$_{\sun}$] & 1.15    &\hspace{- 8pt}$\pm$&\hspace{- 8pt}0.10       & 1.15    &\hspace{- 8pt}$\pm$&\hspace{- 8pt}0.10      \\       
Distance                &[pc]         & 113     &\hspace{- 8pt}$\pm$&\hspace{- 8pt}15         & 245     &\hspace{- 8pt}$\pm$&\hspace{- 8pt}30        \\
$U$                     &[km s$^{-1}$]& $ -8.63$&\hspace{- 8pt}$\pm$&\hspace{- 8pt}0.86       &$ -5.61$ &\hspace{- 8pt}$\pm$&\hspace{- 8pt}1.73      \\
$V$                     &[km s$^{-1}$]& $-22.03$&\hspace{- 8pt}$\pm$&\hspace{- 8pt}0.37       &$-18.14$ &\hspace{- 8pt}$\pm$&\hspace{- 8pt}0.57      \\
$W$                     &[km s$^{-1}$]& $ -6.74$&\hspace{- 8pt}$\pm$&\hspace{- 8pt}0.84       &$ -7.86$ &\hspace{- 8pt}$\pm$&\hspace{- 8pt}1.93      \\
\hline
\end{tabular}
\tablefoot{
\tablefoottext{a}{From the ROSAT All-Sky Survey Faint Source Catalogue \citep{Voges2000}.}
\tablefoottext{b}{$JHK_{\rm s}$ magnitudes are from 2MASS catalogue \citep{2MASS}.}
\tablefoottext{c}{From ROTFIT code.}
\tablefoottext{d}{From {\it FWHM}--$v\sin i$ calibration.}
\tablefoottext{e}{Calculated by means of a spectral synthesis based on {\tt ATLAS9} atmospheric models (Sect.\,\ref{Sec:Abundance}).}
}
\end{table}


\section{Target characterization from ground-based observations}
\label{Sec:Char}

\subsection{Astrophysical parameters}	
\label{Sec:Analysis}

The analysis of the high-resolution spectra was aimed at measuring the
radial ($RV$) and projected rotational velocities ($v\sin i$), performing the MK spectral classification, and
deriving basic stellar parameters 
like effective temperature ($T_{\rm eff}$), gravity ($\log g$), and metallicity ([Fe/H]). 

We have computed the cross-correlation functions (CCFs) with the IRAF task \textsc{fxcor} adopting spectra of late-F
and G-type slowly-rotating stars (Table\,\ref{Tab:Standards}),
which were acquired with the same setups and  during the same observing nights as our targets. 
The latter were used both for the measure of $RV$ and the determination of $v\sin i$ through a calibration of 
the full-width at half maximum ({\it FWHM}) of the CCF peak as a function of the $v\sin i$ of their artificially 
broadened spectra \citep[see, e.g.,][]{Guillout09,Martinez10}. 
We averaged the results from individual {\it \'echelle} orders as described, e.g., in \citet{Frasca2010} to get
the final $RV$ and $v\sin i$ of our young Sun-like stars. 

For KIC\,7985370, we measured, within the errors, the same $RV$ in the FRESCO spectra taken on 2009 July 10 ($RV=-23.9\pm 0.4$~km\,s$^{-1}$) and on 2009 October 
4 ($RV=-24.2\pm 0.4$~km\,s$^{-1}$), while a value of $RV=-21.8\pm 0.2$~km\,s$^{-1}$ has been derived from the SARG spectrum of 2009 August 12. 
This could indicate a true $RV$ variation just above the 3-$\sigma$ confidence level. However, much more spectra are needed to prove or disprove such a $RV$ 
variation and to make sure that the star is a single-lined spectroscopic binary.

For KIC\,7765135, we only acquired one spectrum from which we derived an $RV$ of $-20.0\pm$0.2~km\,s$^{-1}$.


Since we found no indication for binarity (visual or spectroscopic) in the literature for both stars, we consider them as being 
single stars or, at most, single-lined spectroscopic binaries for which the eventual companions are too faint to give a significant contribution
neither to the optical spectrum nor to the \kep photometry. 

For the SARG spectrum, we used the ROTFIT code \citep{Frasca03,Frasca06} to evaluate $T_{\rm eff}$, $\log g$, [Fe/H], and re-determine
$v\sin i$, adopting a library of ELODIE Archive spectra of standard stars, as described in Paper~I. 
To match the lower resolution ($R_{\rm FOCES}=28\,000$) of the FOCES spectrum, the ELODIE templates ($R_{\rm ELODIE}=42\,000$) were 
convolved with a Gaussian kernel of ${\it FWHM} = \lambda\sqrt{1/R_{\rm FOCES}^2-1/R_{\rm ELODIE}^2}$\,\AA\  before running ROTFIT. 
We applied the ROTFIT code only to the {\it \'echelle} orders with a fairly good S/N, which span the ranges 5600--6700\,\AA\  and
4300--6800\,\AA\  for SARG and FOCES spectra, respectively. The spectral regions heavily affected by telluric lines were excluded from the analysis. 
The adopted estimates for the stellar parameters and the associated uncertainties
come from a weighted mean of the values derived for all the individual
orders, as described in Paper~I. 
We also applied this procedure to the FRESCO spectra of  KIC\,7985370, obtaining stellar parameters in 
very close agreement with those derived from the SARG spectrum.


ROTFIT also allowed us to measure $v\sin i$ by matching to the observed spectrum the spectra of slowly-rotating standard stars 
artificially broadened at an increasing $v\sin i$  and finding the minimum of $\chi^2$. 
For this purpose, we used the standard stars listed in Table\,\ref{Tab:Standards} because their spectra were acquired 
with the same instrumental setup as that used for our targets, avoiding the introduction of any systematic error caused by a 
different resolution.

We obtained nearly the same values of the astrophysical parameters for both targets. They turn out to be very similar to the Sun, both in
spectral type and effective temperature, but considerably younger, as testified by the high lithium content in their photospheres
(Sect.\,\ref{Sec:Abundance}). Their average values and standard errors are reported in Table~\ref{Tab:StarChar}.

Based on these parameters and the age estimate (Sects.\,\ref{Sec:Abundance} and \ref{Sec:Kinematics}), compatible with zero-age 
main-sequence (ZAMS) stars, we can roughly estimate their mass and radius as $M=1.15\pm\,0.10\,M_{\sun}$ and  $R=1.1\pm\,0.1\,R_{\sun}$ 
from evolutionary tracks \citep[e.g.,][]{Siess00}. 
 


\subsection{Chromospheric activity}		
\label{Sec:Chrom}

\begin{figure*}[th]
\centering
 \includegraphics[width=17cm]{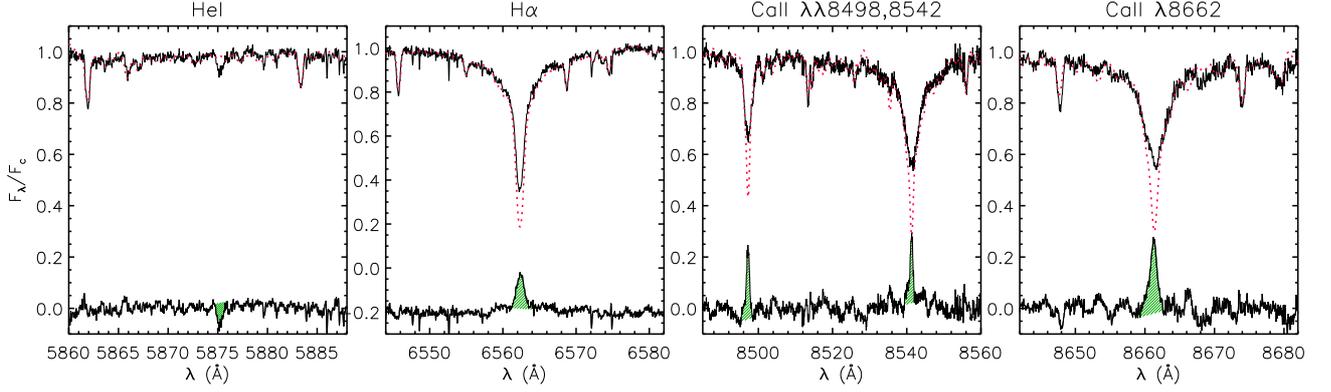}
\caption{{\it Top of each panel}: Observed, continuum-normalized SARG spectrum of KIC\,7985370 
(solid line) in the \ion{He}{i} D$_3$, H$\alpha$, and Ca\,{\sc ii} IRT regions together with the non-active 
stellar template (dotted red line). {\it Bottom of each panel}: Difference between observed and template spectra.
The residual H$\alpha$ profile is plotted shifted downwards by 0.2 for the sake of clarity.
The hatched areas represent the excess emissions (absorption for \ion{He}{i}) that have been integrated to get the net equivalent widths.} 
\label{Fig:Halpha_IRT_79}
\end{figure*}

\begin{figure*}[th]
\centering
 \includegraphics[width=17cm]{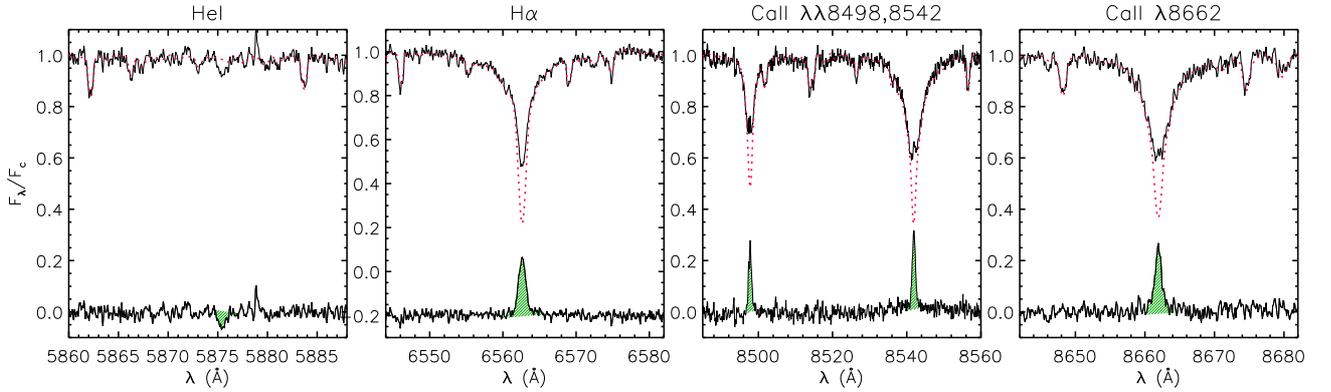}
\caption{As Fig.~\ref{Fig:Halpha_IRT_79}, but for the FOCES spectrum of KIC\,7765135. }
\label{Fig:Halpha_IRT_77}
\end{figure*}

The high level of magnetic activity in the chromosphere is displayed by the \ion{Ca}{ii}
IRT and  H$\alpha$ lines that are strongly filled-in by emission (Figs.\,\ref{Fig:Halpha_IRT_79} and \ref{Fig:Halpha_IRT_77}). 
The \ion{He}{i} D$_3$ ($\lambda\,$5876\AA) 
absorption line is clearly visible in the spectra of our two young suns (Figs.\,\ref{Fig:Halpha_IRT_79} and \ref{Fig:Halpha_IRT_77}). 
This implies the presence of an upper chromosphere 
because a temperature of at least 10\,000\,K is required for its formation. 

We estimated 	
the chromospheric emission level 
with the ``spectral subtraction'' technique \citep[see, e.g.,][]{Frasca94, Montes95}. This procedure 
is based on the subtraction of a reference ``non-active''	
template made with observed spectra of slowly-rotating stars of 
the same spectral type as the investigated active stars, but with a negligible level of chromospheric activity and 
{an undetectable lithium line.} 
The equivalent width of the lithium and helium absorption lines ($W_{\rm Li}$ and $W_{\rm He}$, respectively) and the net equivalent width  
($W^{\rm em}$) of the \ion{Ca}{ii} IRT and H$\alpha$ lines were measured from the spectrum obtained after 
subtracting the non-active template from the target and then by integrating
the residual emission (absorption for the lithium and helium lines) profile, which is represented by the hatched
areas in Figs.\,\ref{Fig:Halpha_IRT_79} and \ref{Fig:Halpha_IRT_77}.  

We noted a significant variation of the H$\alpha$ emission in the three spectra of KIC\,7985370, with 
$W_{\rm H\alpha}^{\rm em}$ ranging from 160 to 264 \,m\AA\ (Table\,\ref{tab:fluxes}). 

{The radiative losses in the chromospheric lines were evaluated by multiplying the average $W^{\rm em}$ by the continuum 
surface flux at the wavelength of the line, in the same way as in \citet{Frasca2010}. }
%
The net equivalent widths and the chromospheric {fluxes} are reported in Table~\ref{tab:fluxes}.

The H$\beta$ net equivalent width could be measured only for KIC\,7765135 because the SARG spectrum of KIC\,7985370 does not
include that wavelength region. In the FRESCO spectra, it is not possible to apply the subtraction technique, due
to the very low S/N ratio in that spectral region, which prevented us from detecting such a tiny filling of the line.

For KIC\,7765135 we measured a Balmer decrement $F_{\rm H\alpha}/F_{\rm H\beta} = 4.3\pm 2.5$ that is slightly larger than the values
in the range 2--3 found by us for \object{KIC~8429280} (Paper~I) 
and for \object{HD~171488}  \citep{Frasca2010}.
Accounting for the error, this ratio is however still compatible with optically-thick emission by atmospheric features similar to solar 
and stellar plages \citep[e.g., ][]{Buza89,Chester91} rather than to prominence-like structures, for which a Balmer decrement of the 
order of 10 is expected \citep[e.g., ][]{LandMong79,Hall92}.

{The \ion{Ca}{ii}-IRT flux ratio,} $F_{8542}/F_{8498}=1.4\pm 0.9$, is indicative of high optical depths and is in the 
range of the values found by \citet{Chester91} in solar plages. 
{The optically-thin emission from solar prominences gives rise instead to values of $F_{8542}/F_{8498}\sim 9$. 
We measured a value of $F_{8542}/F_{8498}=1.6\pm 1.0$ for KIC\,7985370, that is almost the same as KIC\,7765135.}


{The chromospheric emission in both stars appears thus to be mainly due to surface regions similar to solar plages, while
the quiet chromosphere and eventual prominences play a marginal role.}

\begin{table}
\caption{Line equivalent widths and {chromospheric fluxes.}}
\label{tab:fluxes}
\centering
 \begin{tabular}{lccccc}
  \hline\hline
  \noalign{\smallskip}
  Line                  & Date  & UT$_{\rm mid}$  & $W^{\rm em}$  &  Error  & Flux                \\  			     
                        & \scriptsize{(yyyy/mm/dd)}  & \scriptsize{(hh:mm)}  & (m\AA)	  & (m\AA)   &\scriptsize{(erg\,cm$^{-2}$\,s$^{-1}$)} \\
  \noalign{\smallskip}
  \hline
  \noalign{\smallskip}
  \multicolumn{6}{c}{KIC\,7985370} \\
  \noalign{\smallskip}
  H$\alpha$                    & 2009/07/10   & 22:29  & 160    & 40 & 1.23$\times10^6$    \\
    "	                       & 2009/08/12   & 04:18  & 264    & 30 & 2.02$\times10^6$    \\
    "	                       & 2009/10/03   & 19:17  & 202    & 35 & 1.55$\times10^6$    \\
  \ion{He}{i} D$_3$            & 2009/08/12   & 04:18  & $-50$\tablefootmark{a} & 20 & ...   \\
  \ion{Ca}{ii} $\lambda8498$   & " ~~~"       & " ~~"  & 263 & 55 &  1.33$\times10^6$   \\
  \ion{Ca}{ii} $\lambda8542$   & " ~~~"       & " ~~"  & 439 & 55 &  2.16$\times10^6$   \\
  \ion{Ca}{ii} $\lambda8662$   & " ~~~"       & " ~~"  & 337 & 60 &  1.66$\times10^6$   \\
  \noalign{\smallskip}			    
  \multicolumn{6}{c}{KIC\,7765135} \\
  \noalign{\smallskip}
  H$\alpha$                    & 2009/10/03   & 20:15  & 326   & 60 & 2.48$\times10^6$    \\
  H$\beta$                     & " ~~~"       & " ~~"  & 54   & 30 & 0.58$\times10^6$    \\
  \ion{He}{i} D$_3$            & " ~~~"       & " ~~"  & $-60$\tablefootmark{a} & 25 & ...	     \\
  \ion{Ca}{ii} $\lambda8498$   & " ~~~"       & " ~~"  & 258 & 45 &  1.28$\times10^6$   \\
  \ion{Ca}{ii} $\lambda8542$   & " ~~~"       & " ~~"  & 366 & 45 &  1.76$\times10^6$   \\
  \ion{Ca}{ii} $\lambda8662$   & " ~~~"       & " ~~"  & 328 & 65 &  1.58$\times10^6$   \\
  \noalign{\smallskip}
  \hline
\end{tabular}
\tablefoot{
\tablefoottext{a}{The minus sign for the $W^{\rm em}$ indicates a residual absorption (\ion{He}{i} line).}
}
\end{table}


The subtraction technique also allowed us to measure the lithium equivalent width cleaned up from the contamination of \ion{Fe}{i} 
$\lambda\,6707.4\,$\AA\ line, which is strongly blended with the nearby \ion{Li}{i} line (Fig.~\ref{Fig:lithium}).
The lithium equivalent width measured for KIC\,7985370 ($W_{\rm Li}=155\pm20$\,m\AA) and for KIC\,7765135 ($W_{\rm Li}=160\pm20$\,m\AA), 
just above the Pleiades upper envelope \citep{Soderblom1993}, translates into a very high lithium abundance, $\log N{\rm(Li)} = 3.0\pm0.1$, 
for both stars, as results from the calibrations of \citet{PavMag96}. 
The determination of lithium abundance was refined in Sect.\,\ref{Sec:Abundance} by means of a spectral synthesis based on
{\tt ATLAS9} atmospheric models \citep{kur93}.

\subsection{Abundance analysis}
\label{Sec:Abundance}

\begin{figure}[th]
\centering
\hspace{-0.4cm}
 \includegraphics[width=9cm]{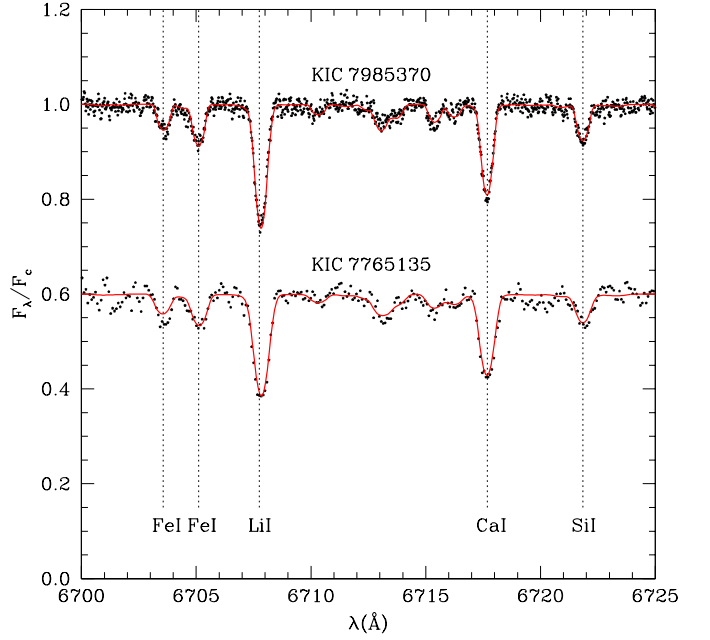}
\caption{Observed spectra (dots) of KIC\,7985370 and KIC\,7765135 (shifted downwards by 0.4) in the \ion{Li}{i} $\lambda\,6707.8\,$\AA\  region
together with the synthetic spectra (full lines). }
\label{Fig:lithium}
\end{figure}

Since the spectral features in our two stars are relatively broad, it is
difficult to find unblended lines for measuring their equivalent widths. 
To overcome this problem, the photospheric abundances
were thus estimated by matching a rotationally-broadened synthetic spectrum to the 
observed one. For this purpose, we divided each spectrum into  50~{\AA}-wide 
segments, which were analyzed separately. 
 The spectral ranges 4450--8650\,\AA\  and 5600--7300\,\AA\  were used for the spectra of KIC\,7765135
 and KIC\,7985370, respectively. 
 The synthetic line profiles were computed with {\tt SYNTHE}
\citep{kur81} using {\tt ATLAS9} \citep{kur93} atmospheric models.  All models 
were calculated using the solar opacity distribution function and a microturbulence 
velocity $\xi$\,=\,2~km~s$^{-1}$.  
For each segment, the abundance was determined by $\chi^2$ minimization.  We used the spectral
line list and atomic parameters from \citet{castelli04} that are updated
from \citet{kur95}.

The error in the abundance of one particular element was taken as the standard 
deviation of the mean of the abundances calculated for each segment.
For elements whose lines occurred in only one or two segments, the error in the 
abundance was evaluated by varying the effective temperature and gravity within 
their uncertainties (as given in Table~\ref{Tab:StarChar}) and computing the abundance 
for $T_{\rm eff}$ and $\log g$ values within these ranges.  We found a variation of 
$\approx$\,0.1~dex in abundance due to temperature variation, but no
significant variation when $\log g$ was varied. Thus the uncertainty in temperature
is probably the main source of error in the abundance estimation. In Fig.~\ref{Fig:lithium} we show an example 
of the matching between synthetic and observed spectra. To determine the 
Li abundance, we used the \ion{Li}{i} $\lambda$6707 {\AA} line, taking into 
account the  hyperfine structure \citep{andersen84}. The abundances are given in Table~\ref{abund}.

\begin{figure} 
\centering
 \includegraphics[width=9cm]{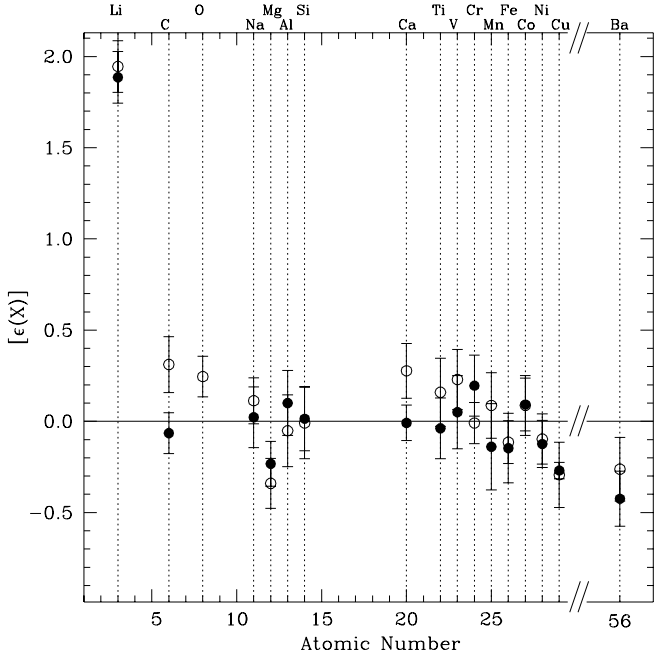} 
\caption{Abundances found for KIC\,7765135 (open circles) and for KIC 7985370 (filled circles) related to the solar ones. } 
\label{patterns} 
\end{figure}

Fig.~\ref{patterns} shows a comparison between the abundances derived for our targets
and the solar values from \citet{grevesse2010}. With the exception of lithium, all the other
abundances are rather similar to those measured for the solar photosphere. 
In particular, we note that KIC\,7765135 is 
$\approx$\,0.3~dex more metallic than the Sun, which confirms the results of ROTFIT (Table~\ref{Tab:StarChar}).

As far as the lithium is concerned, we converted the abundance from $\log N/N_{\rm tot}$ to $\log N/N{\rm(H)}$
on a scale where $\log N{\rm(H)}=12$, finding $\log N{\rm(Li)} = 2.93\pm 0.10$ and $2.87\pm 0.10$ for KIC\,7765135 and  KIC\,7985370,
respectively. These values are smaller than, but marginally consistent with, those obtained from the equivalent widths and the 
calibrations of \citet{PavMag96} and reported in Sect.\,\ref{Sec:Chrom}. 

For early G-type stars, like our targets, the lithium abundance does not give a very strong constraint on their age.
Indeed, their $\log N{\rm(Li)}$ remains close to the initial lithium abundance for Population\,I stars ($\log N{\rm(Li)}$=3.1--3.3) 
during their early life. It has been shown that stars with temperatures in the range $T_{\rm eff}=5900\pm150$\,K display a significant lithium 
depletion ($\log N{\rm(Li)} = 2.9$) only after about 150\,Myr \citep[see, e.g.,][]{Sestito05}. Thus, we can estimate  an age of the order of 
100--200\,Myr for both stars with a lower limit of roughly 30--50\,Myr, being the lithium abundance below the upper envelope 
of the $\alpha$\,Per (age $\approx$ 50\,Myr) and IC\,2602 (age $\approx$ 30\,Myr) clusters \citep{Montes01b,Sestito05}. 
Moreover, the absence of a strong near-IR excess in their spectral energy distributions (see Sect.\,\ref{Sec:SED}) and of accretion signatures 
in the spectrum, allows us to exclude an age of a few million years for both stars that should already be in a post-T~Tauri phase
(age $\geq$\,10--20\,Myr).

\begin{table} 
\caption{Abundances inferred for our two stars expressed in the form $\log N/N_{\rm tot}$.} 
\label{abund} 
\centering
\begin{tabular}{lrr}  
\hline 
\hline 
Elem. & KIC\,7985370 & KIC\,7765135 \\
\hline                            
\noalign{\smallskip}              
Li  &  $-$9.10 $\pm$ 0.10  &  $-$9.04 $\pm$ 0.10    \\ 
C   &  $-$3.67 $\pm$ 0.10  &  $-$3.29 $\pm$ 0.14    \\
O   &      ...~~~~~~~~     &  $-$3.10 $\pm$ 0.10    \\
Na  &  $-$5.77 $\pm$ 0.16  &  $-$5.68 $\pm$ 0.12    \\
Mg  &  $-$4.67 $\pm$ 0.11  &  $-$4.77 $\pm$ 0.13    \\
Al  &  $-$5.48 $\pm$ 0.17  &  $-$5.64 $\pm$ 0.19    \\
Si  &  $-$4.51 $\pm$ 0.17  &  $-$4.48 $\pm$ 0.14    \\
Ca  &  $-$5.70 $\pm$ 0.09  &  $-$5.42 $\pm$ 0.14    \\
Ti  &  $-$7.12 $\pm$ 0.16  &  $-$6.88 $\pm$ 0.15    \\
V   &  $-$8.00 $\pm$ 0.16  &  $-$7.87 $\pm$ 0.14    \\
Cr  &  $-$6.20 $\pm$ 0.16  &  $-$6.40 $\pm$ 0.10    \\
Mn  &  $-$6.74 $\pm$ 0.20  &  $-$6.52 $\pm$ 0.17    \\
Fe  &  $-$4.68 $\pm$ 0.18  &  $-$4.65 $\pm$ 0.11    \\
Co  &  $-$6.95 $\pm$ 0.13  &  $-$6.82 $\pm$ 0.09    \\
Ni  &  $-$5.94 $\pm$ 0.12  &  $-$5.92 $\pm$ 0.12    \\
Cu  &  $-$8.11 $\pm$ 0.06  &  $-$8.13 $\pm$ 0.17    \\
Ba  &  $-$10.28 $\pm$ 0.12 &  $-$10.11 $\pm$ 0.15    \\
\noalign{\smallskip}  
\hline 
\end{tabular}  
\end{table} 

\subsection{Spectral energy distributions}
\label{Sec:SED}
%

\begin{figure}
 \hspace{0.cm}\includegraphics[width=9cm]{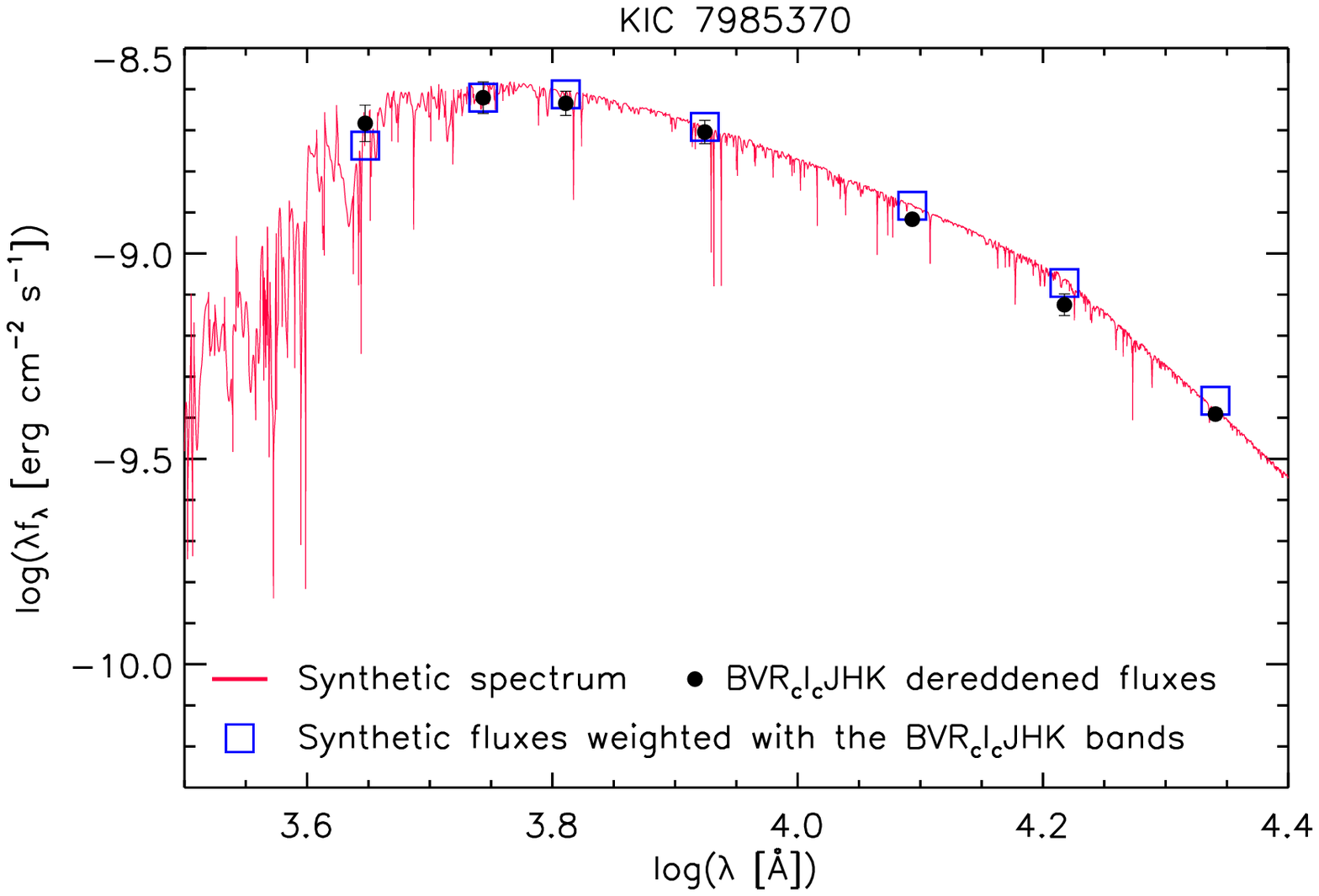}  
 \hspace{0.cm}\includegraphics[width=9cm]{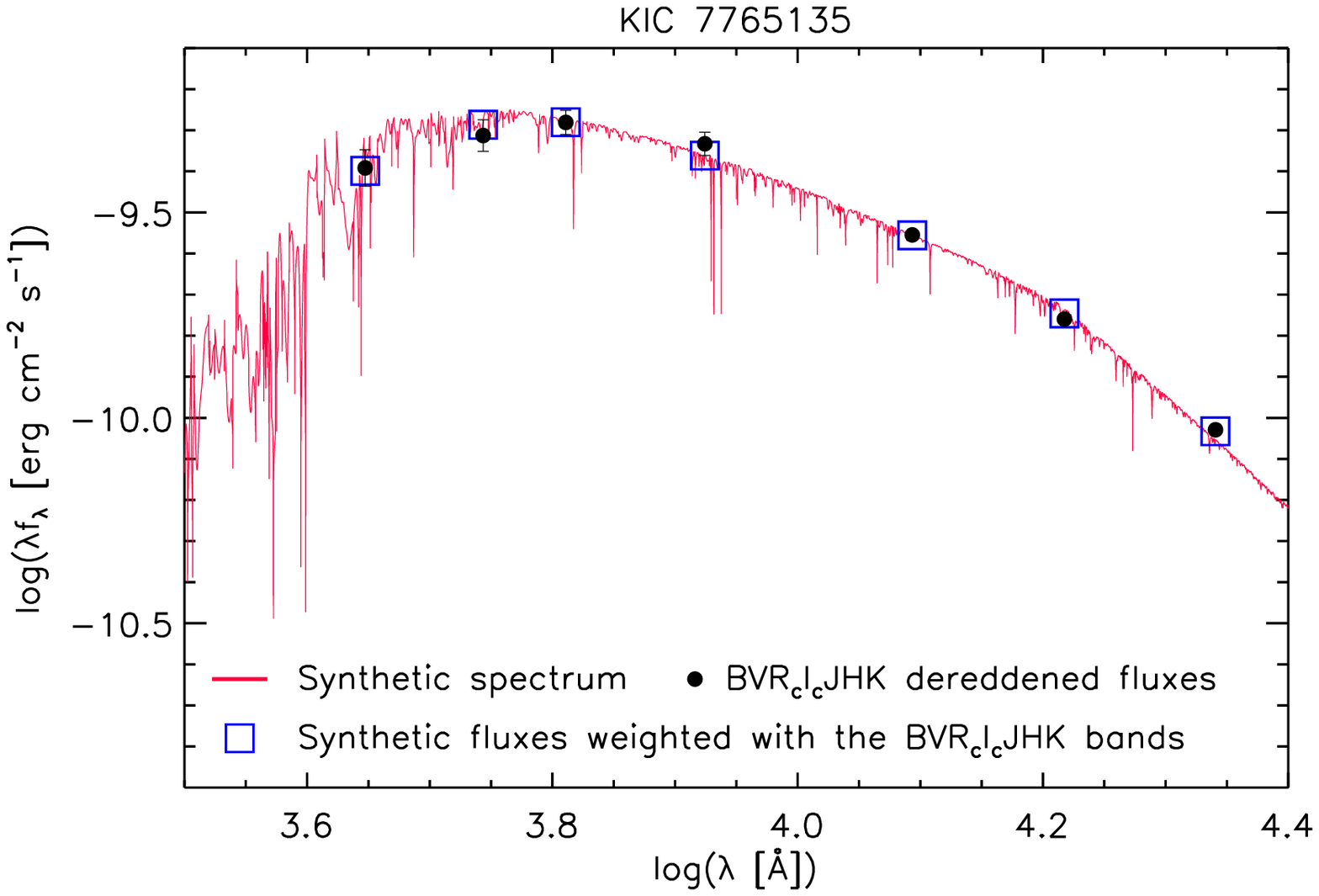}  
\caption{Spectral energy distributions (dots) for KIC\,7985370 {\it (top panel)} and KIC\,7765135 {\it (bottom panel)}. 
The NextGen synthetic spectrum at $T_{\rm eff}=5800$\,K scaled to the star distance is over-plotted with continuous 
lines in each box. }
  \label{Fig:SED}
\end{figure}

\begin{figure*}[ht]
\centering
 \includegraphics[width=17.0cm,height=7.74cm]{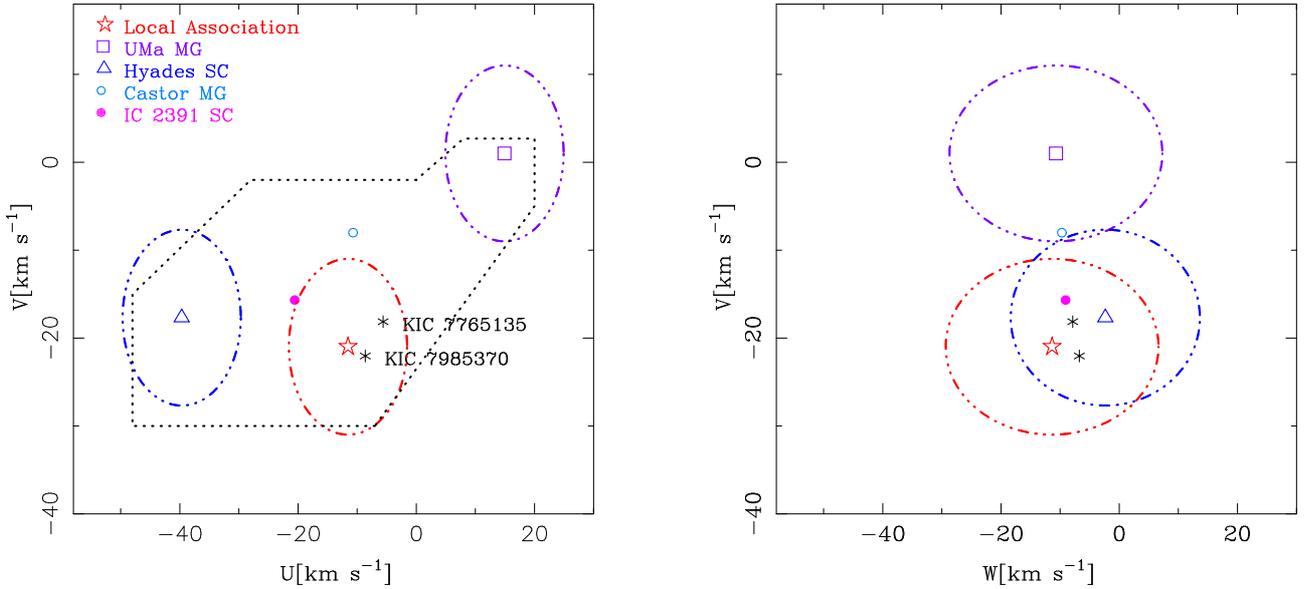}
\caption[ ]{($U$, $V$) and ($W$, $V$) planes of our two young suns. We plotted the average position of each young 
stellar kinematic group 
with different symbols. The dotted line in the left panel demarcates the locus of the young-disk population 
(age $< 2$ Gyr) in the solar neighborhood, as defined by \citet{Eggen84,Eggen89}.
\label{fig:uvw}}
\end{figure*}

The spectral energy distributions (SEDs) from the optical to the near-infrared (IR) domain 
were obtained by merging
our standard $UBVR_{\rm C}I_{\rm C}$ photometry (Table~\ref{Tab:StarChar}) with $JHK_{s}$ magnitudes from the 2MASS catalogue 
\citep{2MASS}. 

We used the grid of NextGen low-resolution synthetic spectra, with $\log g = 4.0$ and 4.5 and solar metallicity by \citet{Hau99a}, to perform a fit 
to the SEDs. The effective temperature ($T_{\rm eff}$) was kept fixed to the value derived with ROTFIT (Table~\ref{Tab:StarChar}) for each target. 
The interstellar extinction ($A_V$) was evaluated from the distance according to the rate of 0.8 mag/kpc found by \citet{Miko} for the sky region around 
CI\,Cyg (very close to our targets). 
{The \citet{Cardelli89} extinction law with $R_V=3.1$ was used for evaluating the extinction in the other bands.}
Finally, the angular diameter ($\phi$) that scales the {synthetic} surface flux over the stellar flux at Earth, was left free to vary.
The best solution was found by minimizing the $\chi^2$ of the fit to the $BVR_{\rm C}I_{\rm C}J$ data, which are dominated by the photospheric 
flux of the star and are normally not significantly affected by infrared excesses. 
{The angular diameter derived for KIC\,7985370,} $\phi=0.0895$\,mas, implies a distance $d=113\pm15$\,pc, if we adopt a ZAMS 
radius of 1.1\,$R_{\sun}$ (Sect.\,\ref{Sec:Analysis}). For KIC\,7765135 we found $\phi=0.0413$\,mas, which corresponds to $d=245\pm30$\,pc, 
adopting again 1.1\,$R_{\sun}$.
The error in the distance is estimated taking into account both a 0.1-$R_{\sun}$ uncertainty in the stellar radius and the temperature error 
of about 100\,K (see Table\,\ref{Tab:StarChar}). 

As apparent in Fig.~\ref{Fig:SED}, the SEDs are {well reproduced} by the synthetic spectrum till the $K_{s}$ band both for KIC\,7985370 and
KIC\,7765135 and no excess is visible at near-IR wavelengths. 
This strengthens the spectroscopic determination of the effective temperature and ensures that these stars have by far gone over the 
T~Tauri phase, during which a thick and dense accretion disk gives rise to a conspicuous infrared excess. 
The absence of mid- and far-infrared data does not allow us to exclude the presence of  thinner ``debris'' disks, like those found in a few
Pleiades solar-type stars \citep{Stauffer2005, Gorlova2006}.

\subsection{Kinematics}
\label{Sec:Kinematics}


We used the radial velocities determined in Sect.~\ref{Sec:Analysis} and the distances estimated in Sect.~\ref{Sec:SED}
together with measurements of proper motions taken from the Tycho-2 catalogue \citep{Hog2000},
to calculate Galactic space-velocity components ($U$, $V$, $W$),
following the procedures described in \citet{Montes01a}.
The values, listed in Table~\ref{Tab:StarChar},
are given in a right-handed coordinate system (positive toward the Galactic anti-center,
in the Galactic rotation direction, and toward the north Galactic pole, respectively).
In the ($U$, $V$) and ($W$, $V$) Bottlinger diagrams (Fig.~\ref{fig:uvw}),
we have plotted the locus of our targets. Different symbols represent the central position
given in the literature \citep[see][]{Montes01a} of the five youngest
and best documented moving groups (MGs) and superclusters (SCs): namely
the Local Association (LA) or Pleiades MG (20--150 Myr),
the IC\,2391 SC (35--55 Myr),
the Castor MG (200 Myr),
the Ursa Major (UMa) MG or Sirius supercluster (300 Myr), and
the Hyades SC (600 Myr).


The space velocities of these two Sun-like stars are consistent with those of the young-disk population (Fig.~\ref{fig:uvw}).
Based on two different statistical methods (for further information, see \citet{Klutschetal2010}
and the paper in preparation by \citet{Klutschetal2012},
we determined their membership probability to each of the five aforementioned young stellar kinematic groups.
KIC~7985370 and KIC~7765135 fall in the  Local Association (LA) locus. With a probability of more than $70$\,\%,
they turn out to be highly likely members of this moving group. Furthermore, such a result is absolutely
in agreement with the age derived from the lithium abundance (Sect.~\ref{Sec:Abundance}).


\section{Spot modelling of the Kepler light curves}
\label{Sec:Spot_mod}

\subsection{Photometric data}\label{data} 



{
All the available public \kep long-cadence
time-series ($\Delta t \approx 30$ min, \citealt{Jenkins2010}),
spanning from 2009 May 2 to 2009 December 16, was analyzed. 
It covers altogether 229 days and corresponds to the observing quarters 0--3 (Q0--Q3), 
with the largest gap, about 4.5 days, appearing between Q1 and Q2.
}


To remove systematic trends in the \kep light curves associated with the spacecraft, detector, and environment, 
and to prepare them for the analysis of star spots that we will describe below, 
we used the software {\sc kepcotrend}\footnote{{\tt http://keplergo.arc.nasa.gov/ ContributedSoftwareKepcotrend.shtml}}.
This procedure is based on Cotrending Basis Vectors (CBV), which are calculated (and ranked) through singular 
value decomposition 
and describe the systematic trends present in the ensemble flux data for each CCD channel. We used the first two basis vectors
for Q0 data, while from three to five CBV were adopted for the correction of longer data sets such as Q1, Q2, and Q3.

In order to check how the data rectification accomplished with {\sc kepcotrend} is reflected in the outcome, 
the spot modelling has been done twice: with the rectified data (Case A) 
as well as with the original data (Case B).

\begin{figure}[h]
 \hspace{-0.4cm}\includegraphics[width=9.5cm]{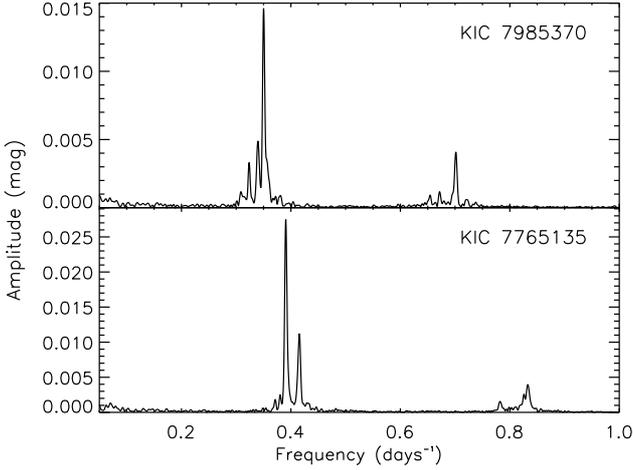}  	
\caption{Cleaned periodograms of the \kep  Q0+Q1+Q2+Q3 time series for KIC\,7985370 (upper panel) and 
KIC\,7765135 (lower panel). } 
  \label{Fig:power}
\end{figure}

The power spectra of the \kep time-series, cleaned by the spectral window according to \citet{Roberts}, are displayed 
in Fig.~\ref{Fig:power}.
The lower panel of Fig.~\ref{Fig:power} clearly shows two main peaks for 
KIC\,7765135, which are close in frequency (0.391 and 0.414 d$^{-1}$).
The corresponding periods are 2.560$\pm$0.015 and 2.407$\pm$0.014 days, respectively.  
{The period errors are from } the {\it FWHM} of the spectral window.
The low-amplitude peaks at frequency of $\approx$\,0.8 d$^{-1}$ are overtones of the two main peaks. 
As visible from the upper panel of Fig.~\ref{Fig:power}, 
the structure of the peaks for KIC\,7985370 is more complex,
with the maximum corresponding to 2.856$\pm$0.019 days
and a second peak, blended with the first one on its low-frequency side, at 2.944 days. A third small peak corresponding to 
$P=3.090$ days is also visible.

{
Such a double- or multiple-peaked periodogram hints at differential rotation. 
As \citet{Lanza94} predicted, a photometric time series, 
if sufficiently accurate (${\Delta F}/{F}=10^{-5}-10^{-6}$), 
may reveal a Sun-like latitudinal differential rotation. 
}

An estimate of the inclination of the rotation axis with respect to the line of sight is very useful to constrain
the spot model. 
{
With $v\sin i$, stellar radius $R$, and rotation period $P$ known, 
the inclination of the rotation axis follows from 
}

\begin{equation}
\label{Eq:vsini}
\sin i = \frac{(v\sin i)\cdot P}{2\pi R}\,. 
\end{equation}
{\noindent  
{
In the absence of an accurate parallax value 
the stellar radius cannot be derived from the effective temperature and luminosity. 
}
If we adopt the radius for a ZAMS star with the effective temperature of our targets ($T_{\rm eff}=5800$\,K),
$R\approx 1.1\,R_{\sun}$, we get  $\sin i = 0.967$ ($i=75\degr\pm15\degr$) for both stars.}
However, as stated in Sect.\,\ref{Sec:Abundance}, the lithium content cannot provide a firm lower limit for the ages of these stars,
which could also be as young as a few 10 Myr (post-T~Tauri phase). 
Thus, allowing for such a young age, a $T_{\rm eff}=5800$\,K is reached by a star of 1.5\,$M_{\sun}$ at 10\,Myr 
with a radius of about 2\,$R_{\sun}$ 
according to the evolutionary tracks by \citet{Siess00}. 
In this case, an inclination of about $30\degr$ is deduced. 


\subsection{Bayesian photometric imaging}\label{Bayes}

The method is basically that of Paper~I. However, as the
light curves are now significantly longer, the introduction of further
free parameters was inevitable. 
The latitudinal dependence of 
rotation frequency $\Omega(\beta)$ (Eq.\,\ref{Eq:angularvelocity}) now contains a $\sin^4\beta$-term and, more important, 
the prescription for spot area evolution is much more detailed. 
Furthermore, the likelihood function (Eq.\,\ref{Eq:likelihood}) is generalized by 
taking into account an unknown linear trend 
in the data. To tackle the problem of strong correlations 
between some parameters, an essential 
new ingredient is the usage of an orthogonalized parameter 
space, where the steered random walk of the Markov chains is performed. 
For the sake of clarity,  due to these new features, the basics 
of the method are explained in this section, although they can also 
found in Paper~1. 
A full account of the method must be deferred to a forthcoming paper
\citep{Froehlich2012}. 

A light-curve fitting that represents spots as dark and circular regions 
has the advantage of reducing the dimensionality 
of the problem and to promptly provide us with average parameters (area, flux contrast, position,
etc.) for each photospheric active region. 
Of course, there are other techniques, which are based on different assumptions, to reconstruct surface 
features photometrically as the inversion of \kep light curves 
done by, e.\,g., \cite{Brown2011}.

Our aim is to present a low-dimensional spot model, with few spots only, that fits 
reasonably well the data regardless to very low-amplitude details that require a high degree of complexity. 
In a Bayesian context this claim could be even quantified. 
Ideally, one should estimate the so-called {\em evidence\/}, 
the integral over the posterior probability distribution. It would 
provide a measure of the probability of
a $n$-spot model and, therefore, allow one to constrain the
number of spots $n$ that are really needed. 
For numerical reasons we are compelled to resort instead to 
the less demanding Bayesian Information Criterion (BIC) by \cite{Schwarz1978}.
The latter or any other related criterion expresses Occam's razor
in mathematical terms without the need to compute the evidence. 
Unfortunately, we have to
admit that -- due to the unprecedented accuracy of the \kep data --
we could not reach this goal with only seven or nine spots. 
There is obviously more information in the data 
than our most elaborate model is able to account for. 

Dorren's (1987) analytical star-spot model, generalized to
a quadratic limb-darkening law, was used. The two coefficients are taken from 
the tables of \cite{ClaretBloemen2011} for a microturbulence velocity 
of $\xi = 2$\,km\,s$^{-1}$ and are used for both the unperturbed photosphere and the spots. 

Four parameters describe the star as a whole: 
One is the cosine of the inclination angle $i$.  
Three parameters ($A, B$ and $C$) describe the latitudinal dependence 
of the angular velocity. 
{
With $\beta$ being the latitude value, the angular velocity $\Omega$ is parameterized 
by a series expansion using Legendre polynomials: 
}
\begin{eqnarray}
\label{Eq:angularvelocity}
\Omega(\beta) = A &+& 3 B\left(5\sin^2 \beta -1\right)/2+ \nonumber \\
&+&\ C\left(315\sin^4 \beta - 210\sin^2 \beta +15\right)/8\,.
\end{eqnarray}
\noindent
The equatorial angular velocity is 
$\Omega_{\rm eq} = A - 3 B/2 +  15 C/8$ 
and the equator-to-pole differential rotation  
${\rm d}\Omega = 15 B/2 + 105 C/8$. In the case of the equator rotating 
faster than the poles ${\rm d}\Omega$ is negative. In what follows the minus sign is 
suppressed, and only the absolute value $|{\rm d}\Omega|$ is given. Both stars 
are definitely rotating like the Sun. 

As in Paper~1, all star spots have the same intensity $\kappa$ relative to the unspotted photosphere
and are characterized by two position coordinates (latitude and initial longitude) and by their radius. 
All these are free parameters in the model. Other spot parameters are the rotation period, which defines 
the spot longitude at any time and is tied to the latitude via Eq.\,\ref{Eq:angularvelocity}. 
The hemisphere, to which a spot belongs to, is to be found by trial and error. 
Further parameters describe the spot area evolution. 
%
%

As our photometric analysis mainly aims at estimating the level of surface differential rotation,
our focus is on {\em long-lived\/} spots. Longevity of star spots 
is at the heart of our approach. In order to obtain, in view of the 
extraordinary length of the time series, a satisfying fit, 
more freedom has been given to spot area evolution with respect to Paper~1.  
It is now parameterized by up to eight parameters. 

{
Spot area is given in units of the star's cross-section. 
Area evolution is assumed to go basically linearly with time. 
}
The underlying physical reason is that then, at least in the case 
of a decaying spot, the slope of the area--time relation
is somehow related to the turbulent magnetic diffusivity. 
With the aim of enhancing flexibility  and to describe the waxing 
and waning of a spot, three consecutive slope values are 
considered. The time derivative of spot area is then 
a mere step function over time. Step height measures  
the increase/decrease of area per day. 
So, there are six free parameters: three 
slope values, two dates of slope change, and the 
logarithm of spot area at some point of the time series. 
We have done even a little bit more. 
In order to prevent sharp bends in spot area evolution, 
some smoothing is introduced. Each date where the slope changes is 
replaced by a time interval within which the slope is linearly
interpolated between the two adjacent values. 
This makes the second time derivative 
of spot area a mere 
step function of time, described by six parameters. 
To get the integrated area itself as function of time 
two constants of integration enter, thus bringing the number 
of free parameters to describe a spot's area evolution to a total of eight.

{
In addition to the free parameters of the model there 
are derived ones, the marginal distributions of which are of interest.
}
An example is the rotational period. It follows from the longitudes 
of the spot centre at the beginning of the time series and at its end. 


{
All parameters are estimated in a Bayesian manner, i.\,e. 
their mean values as well as the corresponding uncertainties follow 
straightforwardly from the data alone. 
}
{
To maintain a flat prior distribution in parameter space, 
all dimensional parameters like periods or spot radii 
must actually be described by their logarithms. 
}
Only then the posterior probability distribution for a
period will be consistent with that of a frequency and likewise the posterior for a
radius with that of an area, i.\,e. it does not matter whether
one prefers periods to frequencies or radii to areas.


The likelihood function  (Eq.\,\ref{Eq:likelihood}) assumes that the 
measurement errors have a Gaussian distribution in the magnitude domain. 
This is justified as long as the 
signal-to-noise ratio does not vary with changing magnitude, as it is
for our data that span a full variation range of less than 0.1 magnitudes. 
It has the invaluable advantage that the likelihood function can be 
analytically integrated over measurement error $\sigma$, 
offset $c_0$, and linear trend $d_0$. To perform the 
integration over $\sigma$ one has to use 
Jeffreys' $1/\sigma$-prior \citep[cf.][]{KassWassermann96}. 
The resulting {\em mean\/} likelihood depends on spot-modelling parameters 
$p_1\dots{}p_M$ only. It takes into account all possible error 
values, offsets and linear trends. By multiplying it with 
the prior, assumed constant in parameter space, one gets 
the posterior density distribution. 
All interesting quantities, parameter averages and confidence intervals, 
are then obtained by marginalization. 


With the $N$ {magnitude values } $d_i$ measured at times $t_i$,
their standard deviations $\sigma_i$, the model magnitudes 
$f_0(t_i,p_1\dots{}p_M)$, offset $c_0$, and 
trend $d_0$, the likelihood {function is given by }
$$\Lambda\left(\sigma, c_0, d_0, p_1\ldots{}p_M; d_i\right) = $$
\begin{equation}
\label{Eq:likelihood}
\prod_{i=1}^{N}{1\over\sqrt{2\pi}\sigma_i}\exp\left(-{\left(d_i-f_0\left(t_i,p_{1\ldots{}M}\right)
  - c_0 - d_0\cdot \left(t_i-t_0\right)\right)^2\over 2\sigma_i^2}\right)\,.
\end{equation}
\noindent
We set $\sigma_i = s_i\cdot\sigma$, 
with relative errors $s_i$ being normalized according to 
$\sum_{i=1}^N{1/s_i^2} = N$.


Parameter estimation by sampling the parameter space has been done 
by the Markov chain Monte Car\-lo (MCMC) method \citep[cf.][]{Press2007}.


Often parameter values are highly correlated. 
As MCMC performs best in an orthogonalized parameter space, 
all parameters have been converted by a principal 
component analysis using singular value decomposition \citep[cf.][]{Press2007}. 
Each parameter in this abstract space is linearly dependent on 
all of the original parameters. The reconstruction of the 
original parameter values can be done exploiting a subspace of that 
orthogonalized parameter space. The dimension of that subspace, 
the number of degrees of freedom, proves lower by roughly one 
third or even more than the number of original parameters. 

\subsection{Results}
\subsubsection{KIC\,7985370}\label{Results:spots:7985370}

\begin{figure*}
\centering
 \includegraphics[width=17cm]{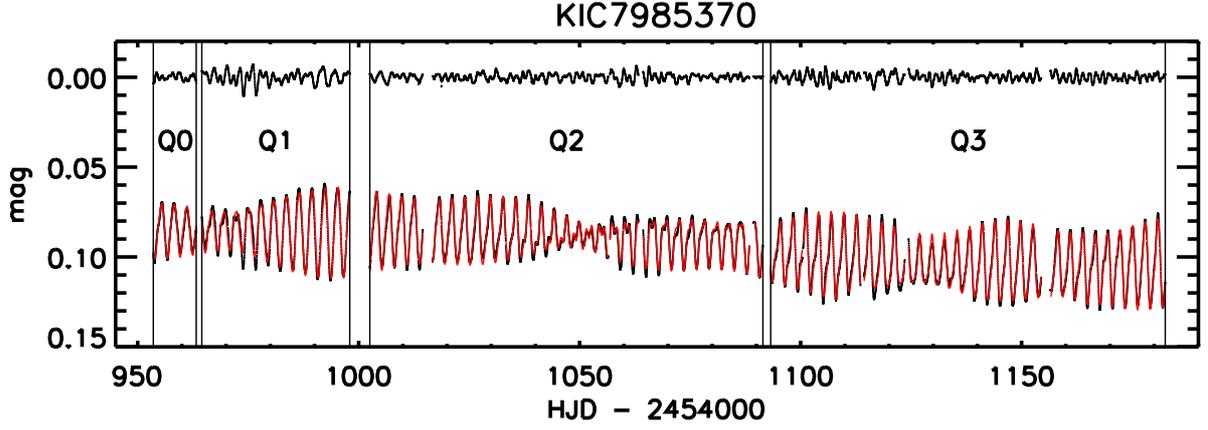}
\vspace{-65mm}
\caption{
{
\kep light curve with best fit (solid red line, 2nd Case-A solution of Table~\ref{7985370Tab}) over-plotted. The residuals, 
}
shown at the top, are $\pm 2.14$\,mmag.
Obviously, the residuals are not homogeneous from one part of the light curve to another. 
}
\label{7985370LightCurve}
\end{figure*}

We have identified eleven gaps longer than an hour and two 
additional small jumps in the light curve (Fig.~\ref{7985370LightCurve}). 
The data set was accordingly divided into 14 parts. Each part 
has been assigned its individual error level, off-set and, in Case B 
(i.\,e. non-rectified data), linear trend. Hence, 
the likelihood (Eq.\,\ref{Eq:likelihood}) is the product 
of 14 independent contributions. 

\begin{figure}[h]
 \resizebox{\hsize}{!}{\includegraphics{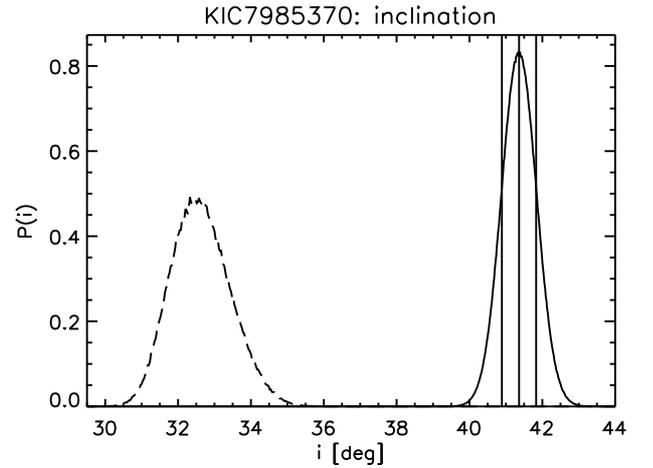}}
\caption{
 Determination of the stellar inclination from \kep photometry. 
 {Mean and 68-per-cent confidence region are marked by vertical lines (Case A only). }
 Dashed: The corresponding marginal distribution for the original data 
 with linear trends removed (Case B).
}
\label{7985370Incl}
\end{figure}

\begin{figure}[h]
  \resizebox{\hsize}{!}{\includegraphics{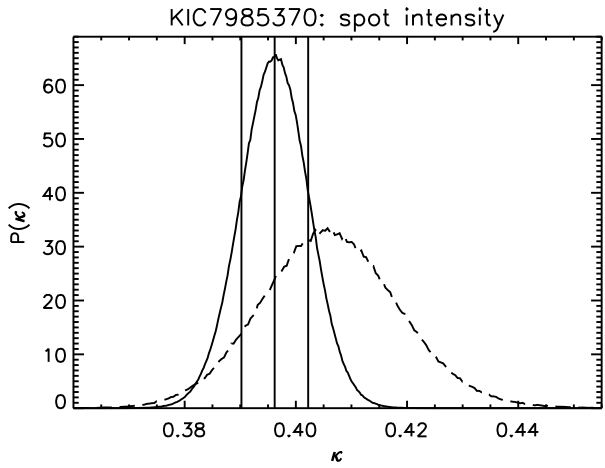}}
\caption{
 Spot intensity {related to } the unspotted photosphere.
 {Mean and 68-per-cent confidence region are marked by vertical lines (Case A only). }
 Dashed: The corresponding marginal distribution 
 for the original data with linear trends removed (Case B).
}
\label{7985370Kappa}
\end{figure}

KIC\,7985370's inclination value $i$ is -- to be honest -- 
ill-defined by the spot model applied to the \kep photometry.
Indeed, with only six spots the MCMC results in very dark spots ($\kappa\approx 0$) 
at very low inclination ($i\approx 10^\circ$). 
But even for these unrealistic solutions 
the equator-to-pole differential rotations was 0.18 rad\,d$^{-1}$. 
Only with seven spots and allowing for enough spot evolution 
we arrived at acceptable inclination values (Fig.~\ref{7985370Incl}) 
and spot intensities (Fig.~\ref{7985370Kappa}). 
If the inclination is fixed to the spectroscopically derived value of $i = 75^\circ$ 
the residuals are rather high, $\pm 2.46$ mmag, exceeding the residuals of 
our best solution ($\pm 2.14$ mmag) by far. Nevertheless, details of the solution with 
fixed inclination are also included in Table~\ref{7985370Tab}, where the results are presented. 

Improving the 7-spot solution by adding an eighth spot leads formally to a better fit. 
As the new spot proves to be ephemeral, lasting only six rotations, it neither 
constrains the differential rotation nor adds any significant insight
(one can always get a better result by adding short-lived features).

%


\begin{figure}[h]
 \resizebox{\hsize}{!}{\includegraphics{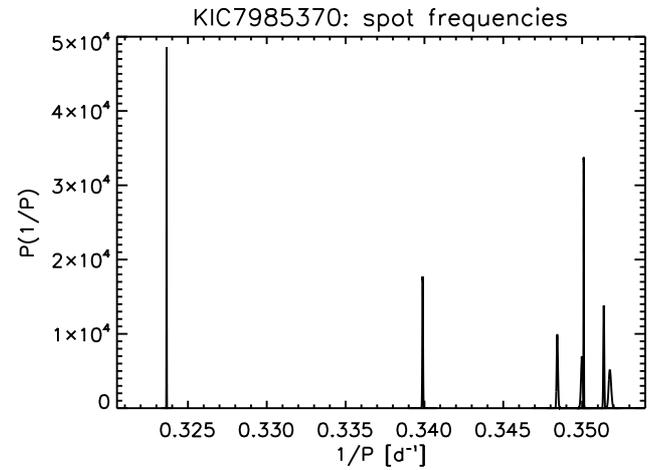}}
\caption{
All seven marginal distributions (Case A) of the spot frequencies. 
The three frequencies (0.324, 0.340, and 0.350 d$^{-1}$) seen 
in the low-resolution Fourier spectrum (Fig.~\ref{Fig:power}) are confirmed 
by the results of our spot model.  
}
\label{7985370Periods}
\end{figure}

The marginal distributions of the seven spot frequencies (Case A only), 
combined into one plot, 
are shown  in Fig.~\ref{7985370Periods}. 

\begin{figure}[h]
 \resizebox{\hsize}{!}{\includegraphics{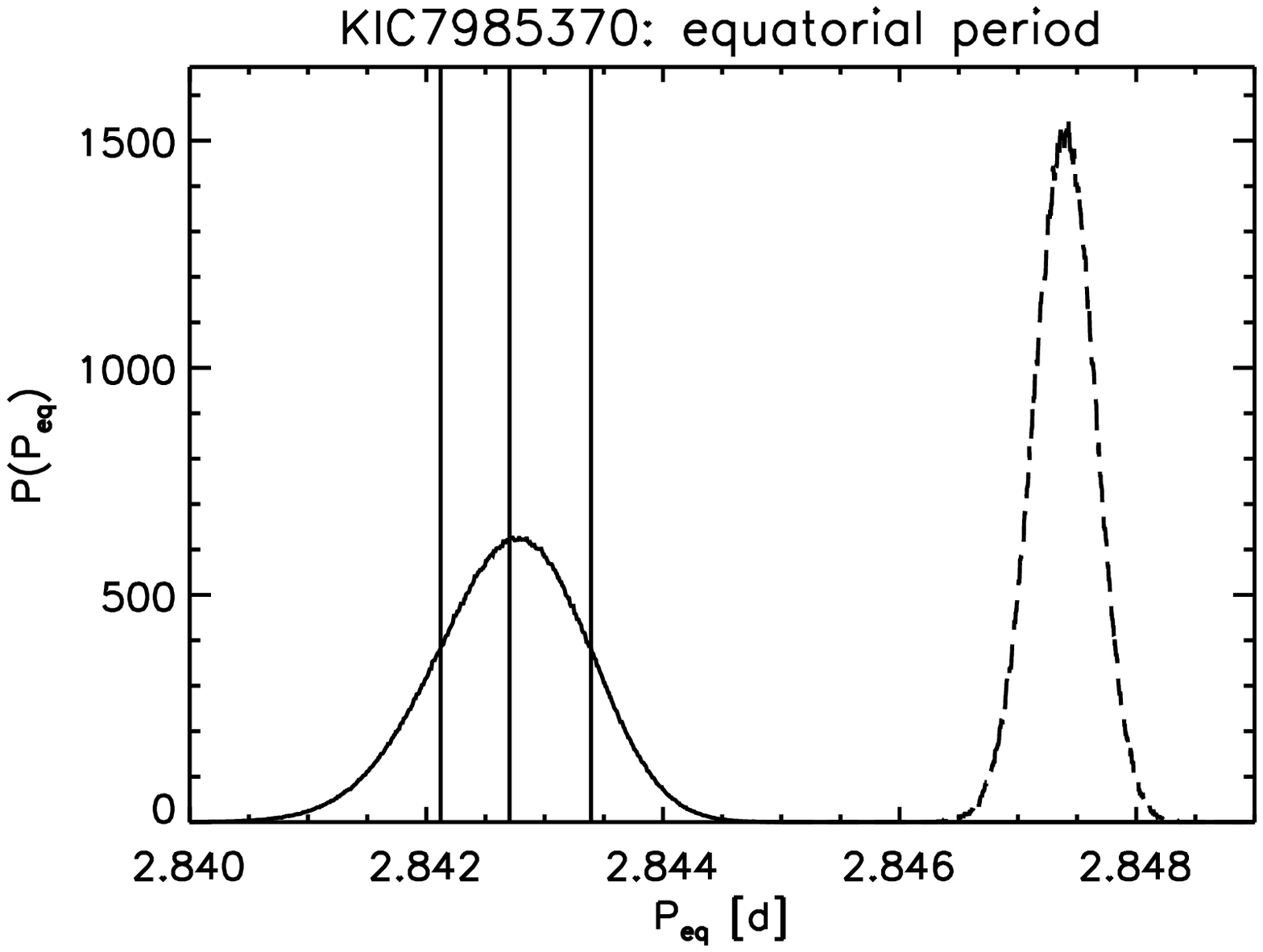}}
\caption{
 Equatorial period of the star. 
 {Mean and 68-per-cent confidence region are marked by vertical lines (Case A only). }
 Dashed: The corresponding marginal distribution for the original data 
 with linear trends removed (Case B).
}
\label{7985370EquPeriod}
\end{figure}

\begin{figure}[h]
 \resizebox{\hsize}{!}{\includegraphics{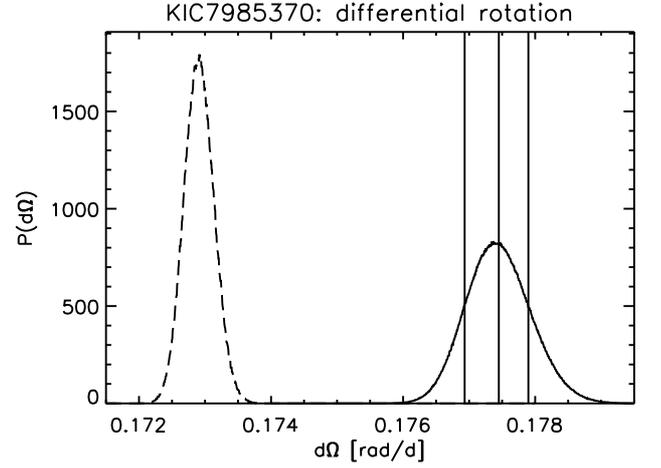}}
\caption{
 Equator-to-pole differential rotation of the star. 
 {Mean and 68-per-cent confidence region are marked by vertical lines (Case A only). }
 Dashed: The corresponding marginal distribution for the original data 
 with linear trends removed (Case B).
}
\label{7985370DiffRot}
\end{figure}

From the three parameters describing the star's surface rotation, 
$A, B$ and $C$, the equatorial rotational period (Fig.~\ref{7985370EquPeriod}) 
and the equator-to-pole differential rotation (Fig.~\ref{7985370DiffRot}) 
follow. The latter amounts to $0.1774^{+0.0004}_{-0.0005}$ rad\,d$^{-1}$ (Case A) 
and $0.1729\pm 0.0002$ rad\,d$^{-1}$ (Case B), respectively. The 
difference is significant, considering the formal errors, albeit very small.
In the case of fixed inclination ($i = 75^\circ$) the differential rotation 
would be slightly enhanced, $0.1839\pm 0.0002$ rad\,d$^{-1}$. 

\begin{figure}[h]
 \resizebox{\hsize}{!}{\includegraphics{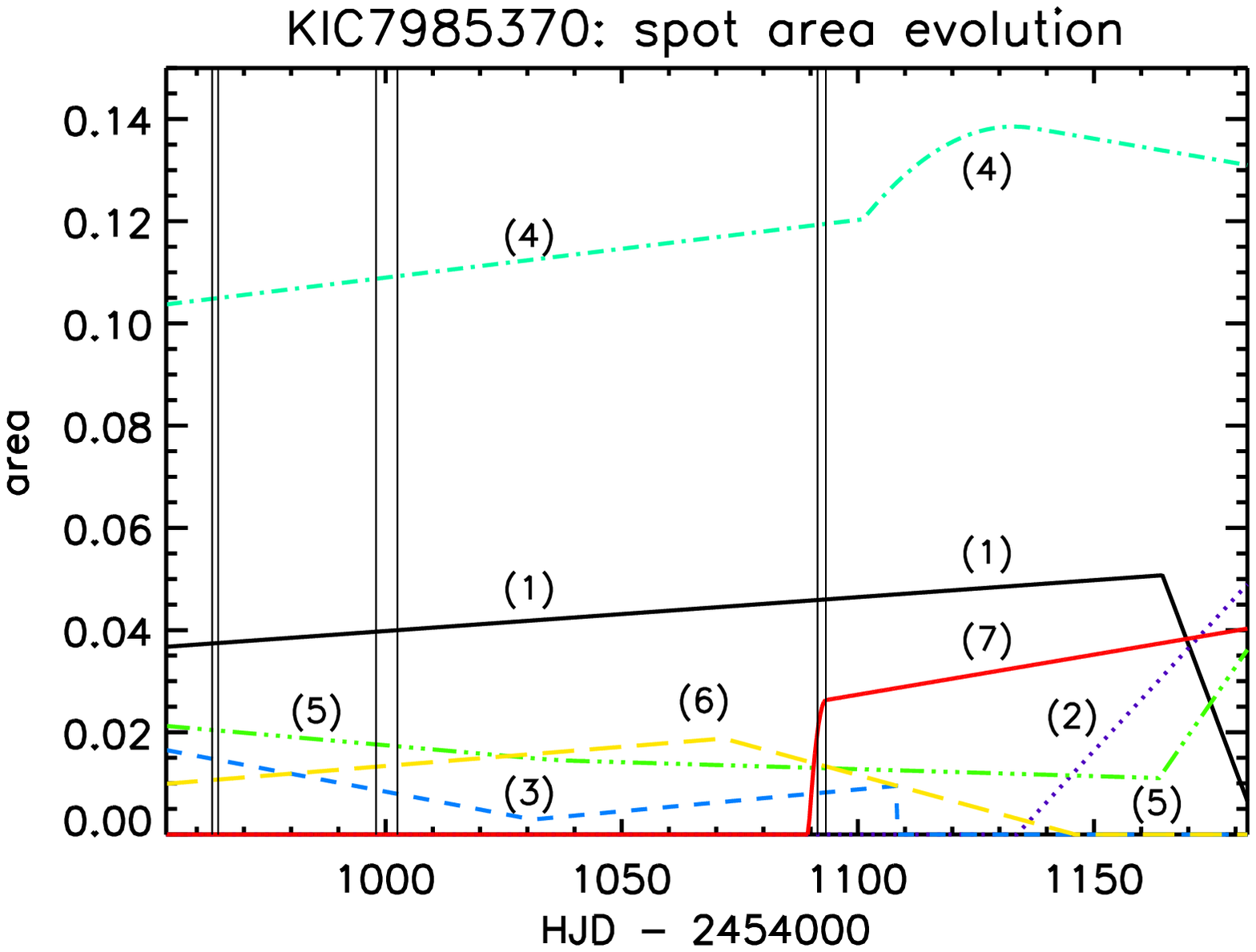}}
\caption{
 Spot area evolution (Case A). Area is in units of the star's cross-section. 
 Vertical lines mark the boundaries of the Q0 to Q3 quarters of data. 
 A number in parenthesis indicates the spot number. 
}
\label{7985370SpotAreaEvolution}
\end{figure}

The spot area evolution is depicted in Fig.~\ref{7985370SpotAreaEvolution}. 
The sudden rise of spot \#\,7 seems to be an artefact. It falls 
into the gap between the end of Q2 data and the beginning of Q3 data. 
On the other hand, the sudden disappearance of spot \#\,3 is not 
related to any switching from one part of the light curve to the next one.
The fall in area of spot \#\,1 at the end of the 
time series is somehow mirrored in an increase in the size of 
spot \#\,5. Maybe this indicates a flaw due to too much 
freedom in describing spot area evolution. 

Expectation values with 1--$\sigma$ confidence limits for
various parameters are also quoted in Table\,\ref{7985370Tab}.

\begin{table*}[ht]
\centering
\caption{Three 7-spot solutions for KIC\,7985370. Listed are {\em expectation\/} values
and 1-$\sigma$ confidence limits.
}\label{7985370Tab}

\begin{tabular}{lllcrlrlrl}
\hline
\hline
\multicolumn{3}{c}{parameter\tablefootmark{a}} & &\multicolumn{2}{c}{Case A\tablefootmark{b}} &\multicolumn{2}{c}{Case A\tablefootmark{b}} &\multicolumn{2}{c}{Case B\tablefootmark{b}}\\
\hline
&&&&&&&\\[-8pt]
\multicolumn{2}{l}{inclination}&$i$&&$ 75^\circ\hspace{-3pt}.0$ & {\em fixed}             &$ 41^\circ\hspace{-3pt}.4$ &$^{+0.5}_{-0.5}$         &$ 32^\circ\hspace{-3pt}.6$ &$^{+0.7}_{-0.9}$        \\ [4pt]
1st&    latitude        &$\beta_1$   &&$ 34^\circ\hspace{-3pt}.0$ &$^{+0.1}_{-0.1}$         &$ 29^\circ\hspace{-3pt}.4$ &$^{+0.4}_{-0.4}$         &$ 22^\circ\hspace{-3pt}.6$ &$^{+0.5}_{-0.6}$        \\ [4pt]
2nd&    latitude        &$\beta_2$   &&$-10^\circ\hspace{-3pt}.0$ &$^{+0.6}_{-0.9}$         &$ -6^\circ\hspace{-3pt}.9$ &$^{+0.9}_{-0.9}$         &$  3^\circ\hspace{-3pt}.3$\tablefootmark{c} &$^{+0.1}_{-0.1}$        \\ [4pt]
3rd&    latitude        &$\beta_3$   &&$ 32^\circ\hspace{-3pt}.2$ &$^{+0.2}_{-0.2}$         &$ 29^\circ\hspace{-3pt}.9$ &$^{+0.3}_{-0.3}$         &$ 27^\circ\hspace{-3pt}.7$ &$^{+0.4}_{-0.4}$        \\ [4pt]
4th&    latitude        &$\beta_4$   &&$ 86^\circ\hspace{-3pt}.8$ &$^{+0.1}_{-0.1}$         &$ 87^\circ\hspace{-3pt}.5$ &$^{+0.1}_{-0.1}$         &$ 87^\circ\hspace{-3pt}.8$ &$^{+0.1}_{-0.1}$        \\ [4pt]
5th&    latitude        &$\beta_5$   &&$ 53^\circ\hspace{-3pt}.8$ &$^{+0.1}_{-0.1}$         &$ 53^\circ\hspace{-3pt}.6$ &$^{+0.1}_{-0.1}$         &$ 51^\circ\hspace{-3pt}.1$ &$^{+0.2}_{-0.2}$        \\ [4pt]
6th&    latitude        &$\beta_6$   &&$ 35^\circ\hspace{-3pt}.8$ &$^{+0.2}_{-0.2}$         &$ 35^\circ\hspace{-3pt}.8$ &$^{+0.3}_{-0.2}$         &$ 33^\circ\hspace{-3pt}.0$ &$^{+0.3}_{-0.3}$        \\ [4pt]
7th&    latitude        &$\beta_7$   &&$ 29^\circ\hspace{-3pt}.6$ &$^{+0.3}_{-0.2}$         &$ 19^\circ\hspace{-3pt}.9$ &$^{+0.9}_{-0.9}$         &$ 10^\circ\hspace{-3pt}.0$ &$^{+0.5}_{-0.6}$        \\ [4pt]
1st&    period          &$P_1$    &&$2.8581$                  &$^{+0.0001}_{-0.0001}$ &$2.8563$                  &$^{+0.0001}_{-0.0001}$ &$2.8572$                  &$^{+0.0001}_{-0.0001}$\\ [4pt]
2nd&    period          &$P_2$    &&$2.8350$                  &$^{+0.0003}_{-0.0003}$ &$2.8428$                  &$^{+0.0007}_{-0.0006}$ &$2.8475$                  &$^{+0.0002}_{-0.0002}$\\ [4pt]
3rd&    period          &$P_3$    &&$2.8541$                  &$^{+0.0004}_{-0.0004}$ &$2.8572$                  &$^{+0.0004}_{-0.0005}$ &$2.8644$                  &$^{+0.0007}_{-0.0005}$\\ [4pt]
4th&    period          &$P_4$    &&$3.0895$                  &$^{+0.0001}_{-0.0001}$ &$3.0898$                  &$^{+0.0001}_{-0.0001}$ &$3.0888$                  &$^{+0.0001}_{-0.0001}$\\ [4pt]
5th&    period          &$P_5$    &&$2.9382$                  &$^{+0.0002}_{-0.0002}$ &$2.9421$                  &$^{+0.0002}_{-0.0002}$ &$2.9417$                  &$^{+0.0003}_{-0.0002}$\\ [4pt]
6th&    period          &$P_6$    &&$2.8629$                  &$^{+0.0003}_{-0.0004}$ &$2.8700$                  &$^{+0.0004}_{-0.0003}$ &$2.8754$                  &$^{+0.0005}_{-0.0004}$\\ [4pt]
7th&    period          &$P_7$    &&$2.8490$                  &$^{+0.0003}_{-0.0002}$ &$2.8460$                  &$^{+0.0002}_{-0.0002}$ &$2.8487$                  &$^{+0.0002}_{-0.0002}$\\ [4pt]
\multicolumn{2}{l}{spot intensity}&$\kappa$	  &&$0.437$   &$^{+0.005  }_{-0.004  }$ &$0.396$   &$^{+0.006  }_{-0.006  }$ &$0.406  $ 		 &$^{+0.012}  _{-0.012  }$\\ [4pt]
\multicolumn{2}{l}{equ. period}   &$P_{\rm eq}$   &&$2.8347$  &$^{+0.0003 }_{-0.0003 }$ &$2.8427$  &$^{+0.0007 }_{-0.0006 }$ &$2.8474 $ 		 &$^{+0.0003} _{-0.0002 }$\\ [4pt]
\multicolumn{2}{l}{deviation}	  &$C/B$	  &&$0.28  $  & 			&$0.28  $  &			     &$0.21   $ 		 &			  \\ [4pt]
\multicolumn{2}{l}{diff. rotation}&${\rm d}\Omega$ &&$0.1839$  &$^{+0.0002 }_{-0.0002 }$ &$0.1774$  &$^{+0.0004 }_{-0.0005 }$ &$0.1729 $		  &$^{+0.0002} _{-0.0002 }$\\ [4pt]
\multicolumn{2}{l}{residuals}	  &		  &&$\pm 2.46$& 			&$\pm 2.14$&			     &$\pm 2.12$		 &			  \\
\noalign{\smallskip}
\hline
\end{tabular}
\tablefoot{
\tablefoottext{a}{Latitudes $\beta$ are derived from
the assumed law of differential rotation (Eq.\,\ref{Eq:angularvelocity}). Periods $P$ are given in days,
the spot intensity $\kappa$ is in units of the intensity of the unspotted surface.
The ratio $C/B$ measures the deviation from a pure $\sin^2$-law of differential rotation.
The differential rotation ${\rm d}\Omega$ (rad\,d$^{-1}$) is the equator-to-pole
value of the shear. Residuals are in mmag.}
\tablefoottext{b}{Case A refers to rectified data, Case B to non-rectified one.
In order to get the Case-B solution the Case-A solution has been taken
as a starting point for the MCMC parameter estimation.
In the first Case-A solution the inclination is fixed to $i = 75^\circ$.}
\tablefoottext{c}{The second spot is near the equator, therefore, the hemisphere it belongs to is doubtful.}
}
\end{table*}
  

One should be aware that there is more than one solution for each case. 
The second Case-A solution presented in Table\,\ref{7985370Tab} is the 
one that has the lowest residuals found so far. There is an other 
well-relaxed 7-spot solution with slightly larger residuals nearby in parameter space. 
In that solution the fastest spot (\#\,2), coming into existence 
near the end of the time series at JD$\sim$2455135, is 
located at a more southern latitude of $-21^\circ$, 
resulting in a slightly increased differential rotation. 
All other spots are virtually unaffected. 
Further details of this second solution are 
given in Table\,\ref{7985370bTab}.

\begin{table}[t]
\caption{A second pair of 7-spot solutions for KIC\,7985370. }
\label{7985370bTab}
\centering
\begin{tabular}{lllcrlrl}
\hline
\hline
\multicolumn{3}{c}{parameter} & &\multicolumn{2}{c}{Case A} &\multicolumn{2}{c}{Case B}\\
\hline
&&&&&&&\\[-8pt]
\multicolumn{2}{l}{equ. period}   &$P_{\rm eq}$   & &$2.8202$     &$\pm 0.0002$ &$2.8209$   &$\pm 0.0002$\\ [4pt]
\multicolumn{2}{l}{diff. rotation}&${\rm d}\Omega$ & &$0.1943$     &$\pm 0.0002$ &$0.1933$   &$\pm 0.0002$\\ [4pt]
\multicolumn{2}{l}{residuals}     &               & &$\pm 2.20$   &             &$\pm 2.21$ &            \\
\noalign{\smallskip}
\hline
\end{tabular}
\tablefoot{The meaning of the entries is the same as in Table\,\ref{7985370Tab},
i.\,e. periods are in days, the differential rotation in rad\,d$^{-1}$, and the residuals in mmag.}
\end{table}


\subsubsection{KIC\,7765135}\label{Results:spots:7765135}

\begin{figure*}[ht]
\centering
 \includegraphics[width=17cm]{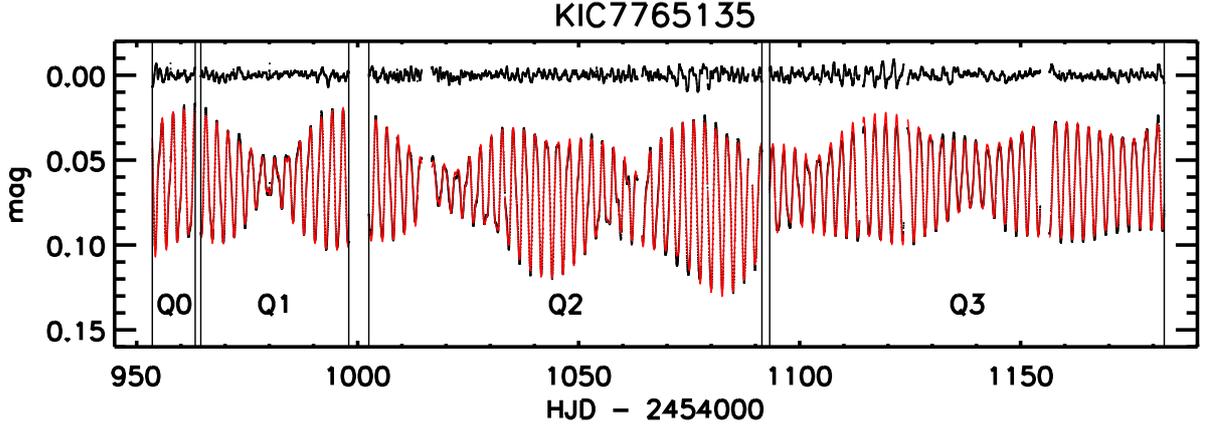}
 \vspace{-65mm}
\caption{
{
\kep light curve with best fit (solid red line, Case-A solution of Table~\ref{7765135Tab}) over-plotted. The residuals, 
}
shown at the top, are $\pm 2.35$\,mmag. 
Obviously, the residuals are not homogeneous from one part of the light curve to another, 
hence, the stated $\pm 2.35$\,mmag is to be considered an overall average. 
}
\label{7765135LightCurve}
\end{figure*}

We have identified eleven gaps longer than an hour and three 
additional small jumps in the light curve (Fig.~\ref{7765135LightCurve}).
The data set was accordingly divided into 15 parts. Each part 
has been assigned its individual error level, off-set and, in Case B
(i.\,e. non-rectified data), linear trend. Hence,
the likelihood function (Eq.\,\ref{Eq:likelihood}) is the product
of 15 independent contributions.

As the inclination is photometrically ill-defined, we fixed it to the 
spectroscopically derived value of $i = 75^\circ$. 

\begin{figure}[h]
 \resizebox{\hsize}{!}{\includegraphics{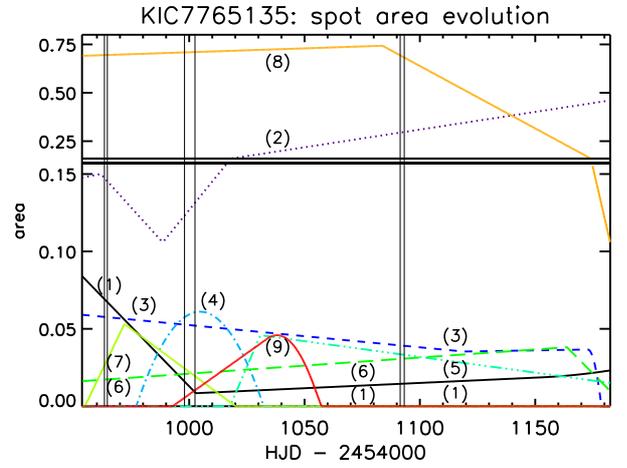}}
\caption{
Same as Fig.~\ref{7985370SpotAreaEvolution}, for KIC\,7765135. 
 Three of the nine spots (\#\,4, \#\,7, and \#\,9) are short-lived ones. 
 Note the change in scale in the upper part! 
}
\label{7765135SpotAreaEvolutionC}
\end{figure}

Despite two spots more, the
residuals, $\pm 2.35$\,mmag, exceed those of the seven-spot model of 
KIC\,7985370 ($\pm 2.14$\,mmag). 
This is not due to the fainter magnitude of KIC\,7765135 compared
to KIC\,7985370, because the photometric uncertainties are typically 0.047\,mmag
for the former and 0.022\,mmag for the latter. 
The reason may be that three of the nine spots are definitely 
short-lived with a life span as low as two months (cf. 
Fig.~\ref{7765135SpotAreaEvolutionC}), which is less than twice 
the lapping time of 38 days  between the fastest and the slowest spot. 
We have to admit that dealing with nine spots 
goes already to the limit of the MCMC technique since the method's relaxation 
time becomes prohibitively long. 


\begin{figure}[h]
 \resizebox{\hsize}{!}{\includegraphics{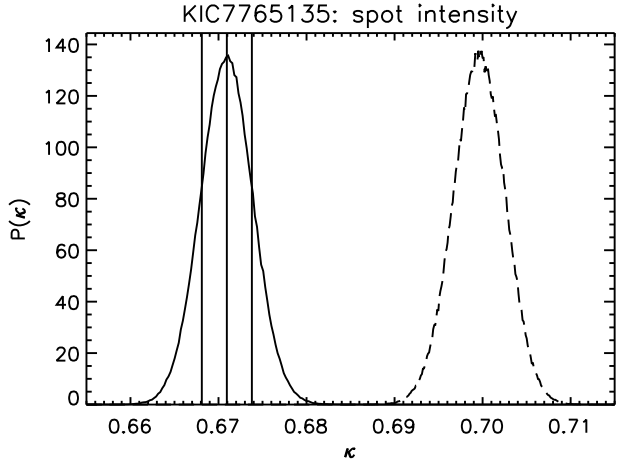}}
\caption{
Same as Fig.~\ref{7985370Kappa}, for KIC\,7765135. 
}
\label{7765135Kappa}
\end{figure}

The marginal distribution of the 
spot rest intensity is shown in Fig.~\ref{7765135Kappa}. 



\begin{figure}[h]
 \resizebox{\hsize}{!}{\includegraphics{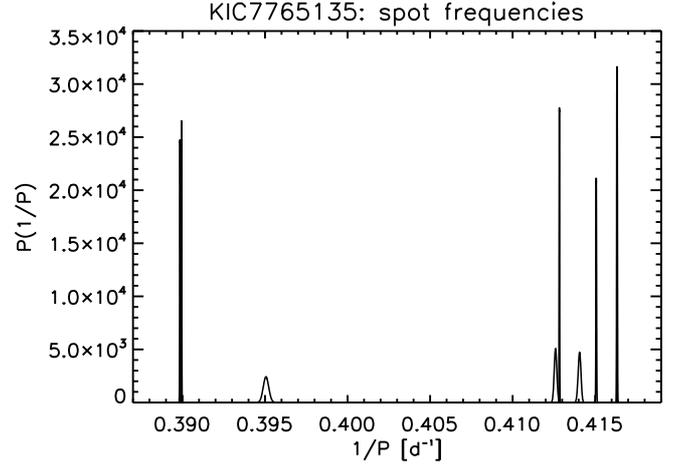}}
\caption{
Marginal distributions of the frequency for all the nine spots. 
The frequency values group around the two principal 
frequencies (0.391 and 0.414 d$^{-1}$) seen already in the 
Fourier spectrum (Fig.~\ref{Fig:power}), which is the reason for the obvious 
``beating'' phenomenon in Fig.~\ref{7765135LightCurve} with a period of 40 days.
The shortest and the longest frequency are a superposition of two frequencies. 
}
\label{7765135Periods}
\end{figure}

The marginal distributions of the nine spot frequencies (Case A only), 
combined into one plot, are shown  in Fig.~\ref{7765135Periods}. 


\begin{figure}[h]
 \resizebox{\hsize}{!}{\includegraphics{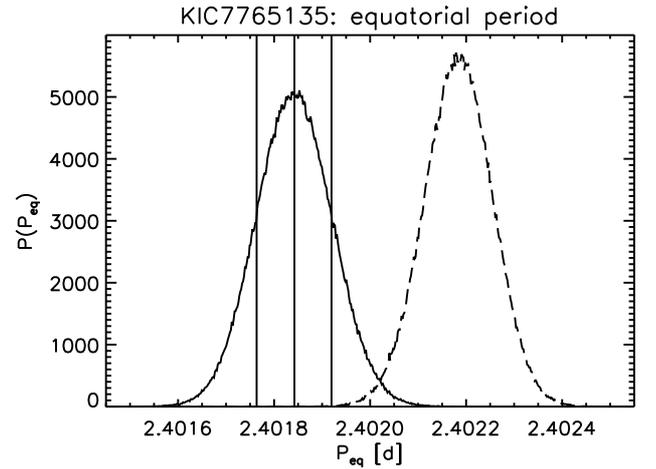}}
\caption{
Same as Fig.~\ref{7985370EquPeriod}, for KIC\,7765135. 
}
\label{7765135EquPeriod}
\end{figure}

\begin{figure}[h]
 \resizebox{\hsize}{!}{\includegraphics{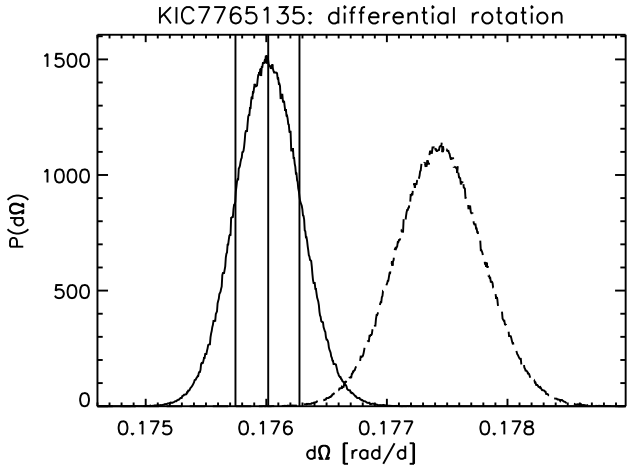}}
\caption{
Same as Fig.~\ref{7985370DiffRot}, for KIC\,7765135. 
}
\label{7765135DiffRot}
\end{figure}

From the three parameters describing the star's surface rotation, 
$A, B$ and $C$, the equatorial rotational period (Fig.~\ref{7765135EquPeriod}) 
and the equator-to-pole differential rotation (Fig.~\ref{7765135DiffRot}) 
follows. The latter amounts to $0.1760\pm 0.0003$ rad\,d$^{-1}$ (Case A)           
and $0.1774^{+0.0003}_{-0.0004}$ rad\,d$^{-1}$ (Case B), respectively. 
As  for KIC\,7985370 the difference is small, but nevertheless significant.

The level of differential rotation does not 
depend on the number of spots considered. Neglecting the three short-lived spots, 
i.e. considering a six-spot model, would result in an equator-to-pole shear of 
$0.1777\pm 0.0006$ rad\,d$^{-1}$. 

Inclination does not significantly affect d$\Omega$. 
Indeed, decreasing the inclination from the adopted value of $i = 75^\circ$ to 
$45^\circ$ would result in a marginally larger
differential rotation, three to four per cent. This is quite understandable.
Inclination affects latitudes, but hardly periods. 

{A cursory glance cast }
at the beating pattern (Fig.~\ref{7765135LightCurve}) 
reveals a lapping time $P_{\rm beat}\sim$\,40 days, which is nearly exactly the lapping time 
of 40.3 days from the two peaks of the cleaned periodogram (Fig.~\ref{Fig:power}). 
From these 40.3 days one already gets an estimate of 
the minimum value for the differential rotation as $2\pi/P_{\rm beat}\sim$\,0.156\,rad\,d$^{-1}$, which is
not far from that derived by the model. 


The spot area evolution (Case A) is depicted in 
Fig.~\ref{7765135SpotAreaEvolutionC}.
The overwhelmingly large southern spot -- at the beginning it fills to a large extent the southern 
hemisphere -- may be an artefact. Because of its southern 
location its contribution to the light curve is rather modest. Perhaps 
it is actually a feature of the northern hemisphere, a non-circular extension of 
spot \#\,2. To prevent spot overlapping, spot \#\,8 had to 
be moved to the southern hemisphere. 
The reader should be aware that even in the case of a large 
spot the whole spot region has been assigned the angular velocity of its 
centre. Differential rotation is, to be exact, not compatible with a
fixed circular shape. This is a shortcoming of our simple model. 

In the case of KIC\,7765135 it cannot be excluded that spot area evolution is partly driven 
by the need to avoid overlapping of spots.

Expectation values with 1--$\sigma$ confidence limits for
various parameters are compiled in  Table\,\ref{7765135Tab}.

\begin{table}[ht]
\caption{Two 9-spot solutions for KIC\,7765135 with inclination being fixed to $i = 75^\circ$.  
Listed are {\em expectation\/} values and 1-$\sigma$ confidence limits. }
\label{7765135Tab}
\centering
\begin{tabular}{lllcrlrl}
\hline
\hline
\multicolumn{3}{c}{parameter\tablefootmark{a}} & &\multicolumn{2}{c}{Case A\tablefootmark{b}} &\multicolumn{2}{c}{Case B\tablefootmark{b}}\\
\hline
&&&&&&&\\[-8pt]
\multicolumn{2}{l}{inclination}&$i$&&$ 75^\circ\hspace{-3pt}.0$ &{\em fixed}            &$ 75^\circ\hspace{-3pt}.0$&{\em fixed}           \\ [4pt]
1st&    latitude        &$\beta_1$   &&$ 20^\circ\hspace{-3pt}.4$ &$^{+0.1}_{-0.1}$       &$ 20^\circ\hspace{-3pt}.0$&$^{+0.1}_{-0.1}$      \\ [4pt]
2nd&    latitude        &$\beta_2$   &&$ 76^\circ\hspace{-3pt}.3$ &$^{+0.2}_{-0.2}$       &$ 75^\circ\hspace{-3pt}.1$&$^{+0.2}_{-0.2}$      \\ [4pt]
3rd&    latitude        &$\beta_3$   &&$  1^\circ\hspace{-3pt}.9$ &$^{+0.2}_{-0.1}$       &$  0^\circ\hspace{-3pt}.8$&$^{+0.2}_{-0.2}$      \\ [4pt]
4th&    latitude        &$\beta_4$   &&$ 21^\circ\hspace{-3pt}.1$ &$^{+0.3}_{-0.3}$       &$ 20^\circ\hspace{-3pt}.5$&$^{+0.3}_{-0.3}$      \\ [4pt]
5th&    latitude        &$\beta_5$   &&$  0^\circ\hspace{-3pt}.9$ &$^{+0.3}_{-0.4}$       &$  0^\circ\hspace{-3pt}.7$&$^{+0.3}_{-0.2}$      \\ [4pt]
6th&    latitude        &$\beta_6$   &&$ 12^\circ\hspace{-3pt}.1$ &$^{+0.1}_{-0.1}$       &$ 11^\circ\hspace{-3pt}.7$&$^{+0.1}_{-0.1}$      \\ [4pt]
7th&    latitude        &$\beta_7$   &&$ 16^\circ\hspace{-3pt}.3$ &$^{+0.3}_{-0.3}$       &$ 17^\circ\hspace{-3pt}.2$&$^{+0.3}_{-0.3}$      \\ [4pt]
8th&    latitude        &$\beta_8$   &&$-75^\circ\hspace{-3pt}.9$ &$^{+0.2}_{-0.2}$       &$-74^\circ\hspace{-3pt}.4$&$^{+0.2}_{-0.2}$      \\ [4pt]
9th&    latitude        &$\beta_9$   &&$ 60^\circ\hspace{-3pt}.2$ &$^{+0.4}_{-0.4}$       &$ 59^\circ\hspace{-3pt}.4$&$^{+0.4}_{-0.4}$      \\ [4pt]
1st&    period          &$P_1$       &&$2.4223$                   &$^{+0.0001}_{-0.0001}$ &$2.4222$                  &$^{+0.0001}_{-0.0001}$\\ [4pt]
2nd&    period          &$P_2$       &&$2.5651$                   &$^{+0.0001}_{-0.0001}$ &$2.5653$                  &$^{+0.0001}_{-0.0001}$\\ [4pt]
3rd&    period          &$P_3$       &&$2.4020$                   &$^{+0.0001}_{-0.0001}$ &$2.4022$                  &$^{+0.0001}_{-0.0001}$\\ [4pt]
4th&    period          &$P_4$       &&$2.4237$                   &$^{+0.0004}_{-0.0004}$ &$2.4231$                  &$^{+0.0005}_{-0.0005}$\\ [4pt]
5th&    period          &$P_5$       &&$2.4019$                   &$^{+0.0001}_{-0.0001}$ &$2.4022$                  &$^{+0.0001}_{-0.0001}$\\ [4pt]
6th&    period          &$P_6$       &&$2.4092$                   &$^{+0.0001}_{-0.0001}$ &$2.4092$                  &$^{+0.0001}_{-0.0001}$\\ [4pt]
7th&    period          &$P_7$       &&$2.4151$                   &$^{+0.0004}_{-0.0005}$ &$2.4172$                  &$^{+0.0005}_{-0.0005}$\\ [4pt]
8th&    period          &$P_8$       &&$2.5645$                   &$^{+0.0001}_{-0.0001}$ &$2.5641$                  &$^{+0.0001}_{-0.0001}$\\ [4pt]
9th&    period          &$P_9$       &&$2.5313$                   &$^{+0.0010}_{-0.0010}$ &$2.5311$                  &$^{+0.0009}_{-0.0009}$\\ [4pt]
\multicolumn{2}{l}{spot intensity} &$\kappa$&&$0.671$           &$^{+0.003  }_{-0.003 }$&$0.700 $                  &$^{+0.003} _{-0.003 }$\\ [4pt]
\multicolumn{2}{l}{equ. period}    &$P_{\rm eq}$   &&$2.4018$   &$^{+0.0001 }_{-0.0001}$&$2.4022$                  &$^{+0.0001}_{-0.0001}$\\ [4pt]
\multicolumn{2}{l}{deviation}      &$C/B$          &&$-0.008$   &                       &$-0.008$                  &                      \\ [4pt]
\multicolumn{2}{l}{diff. rotation} &${\rm d}\Omega$ &&$0.1760$   &$^{+0.0003 }_{-0.0003}$&$0.1774  $                &$^{+0.0003}_{-0.0004}$\\ [4pt]
\multicolumn{2}{l}{residuals}      &               &&$\pm 2.35$ &                       &$\pm 2.29$                &                      \\
\noalign{\smallskip}
\hline
\end{tabular}
\tablefoot{
\tablefoottext{a,b}{The meaning of the superscripts is the same as in Table~\ref{7985370Tab}.}
}
\end{table}

\section{Discussion}
\label{Sec:Disc}

\subsection{Chromospheric and coronal activity}

With the aim of making a comparison with the chromospheric activity of stars similar to our targets, we considered
eighteen stars in the Pleiades cluster, among those investigated by \citet{Soderblom1993b}, that have an early-G spectral type
or an effective temperature close to our targets and a $v\sin i$ between 4 and 18\,km\,s$^{-1}$. With a spectral subtraction analysis they 
found values of the net H$\alpha$ emission between 100 and 500\,m\AA\  and surface fluxes ranging from 1.0$\times10^6$ to 2.9$\times10^6$ 
erg\,cm$^{-2}$\,s$^{-1}$. The net emission filling the core of the \ion{Ca}{ii}-$\lambda8542$ line ranges from about 300 to 600 m\AA\  and
the corresponding flux is in the range 1.0--2.7$\times10^6$ erg\,cm$^{-2}$\,s$^{-1}$. The values of these activity indicators for our two 
targets are inside all these ranges, suggesting an activity level comparable to the Pleiades stars (age $\sim$\,130\,Myr).  
The X-ray luminosity for these Pleiades members, as evaluated by \citet{Stauffer1994} and \citet{Marino03}, is in the range $\log L_{\rm X}=28.7$--29.7, which is 
just below the saturation threshold of $\log L_{\rm X}=30.0$ found by \citet{Pizzolato03} for Sun-like stars. Moreover, according to the
works of \citet{Pizzolato03} and \citet{Soderblom1993b}, the saturation would occur for rotation periods shorter than about 2.0 days. 
Another helpful work dealing with chromospheric activity in stars belonging to the young open clusters IC\,2391 and IC\,2602
(age $\approx 30-50$\,Myr) shows that the chromospheric flux measured in the core of \ion{Ca}{ii}$-\lambda8542$ line saturates
at a value of $\log(R'_{8542})=\log(F'_{8542}/\sigma T{\rm eff}^4) \approx -4.2$ \citep{Marsden09}. A similar behaviour was already found for the 
Pleiades by \citet{Soderblom1993b}. The $\log(R'_{8542})$ values of $-4.5$ and 
$-4.6$ that can be derived for KIC\,7985370 and KIC\,7765135, respectively, are lower than, but not very far from, this saturation level.
From the ROSAT X-ray count and the distance quoted in Table\,\ref{Tab:StarChar} we evaluated the X-ray luminosity of KIC\,7985370 through
the relation proposed by \citet{Fleming1995}, $L_{\rm X}= 4\pi d^{2}(8.31 + 5.30\,HR1)\times 10^{-12}$\,erg\,s$^{-1}$, where the hardness ratio 
is $HR1=0.23$ \citep{Voges2000}.
The value of $\log L_{\rm X}=29.68^{+0.15}_{-0.20}$ confirms the non-saturated regime for KIC\,7985370. 

\subsection{General considerations on the spot model}

Despite the fact that both stars are very active ones and exhibit filled-in absorption of several chromospheric activity indicators, 
our photometric analysis is in terms of dark surface features only. Allowing for bright ones too, would make 
the MCMC approach usually unstable. Anyway, photometry alone seems to be unable to 
discriminate even between dark and bright spots \citep[e.g.,][]{Lueftinger2010}. 

As a comparison of the two cases A and B reveals, 
the results of our spot modelling are hardly influenced 
by the rectification procedure.
The smaller residuals for non-rectified data (Case B) are very likely due to the fact that the Case-B 
likelihood function (Eq.\,\ref{Eq:likelihood}) also takes into 
account a linear trend in the data, individually for each 
part of the light curve. This allows for more freedom in fitting the data and 
results in a slightly better fit.


The reader should be aware that the estimated parameter values and their
often surprisingly small errors 
are those of the model constrained by the data. 
Error bars indicate the ``elbow room'' of the model, nothing more.

\subsection{Frequencies}

It is remarkable that the frequencies that stand out in the power spectrum of the 
light curves (Fig.~\ref{Fig:power}) represent the distribution of spot frequencies (Figs.~\ref{7985370Periods} and \ref{7765135Periods}) 
astonishingly well. The lapping time, as a measure of the lower limit of surface differential rotation, follows already 
from the periodogram analysis! 
However, in order to get an estimate of the full equator-to-pole span of the 
latitudinal shear, including its sign, one needs latitudinal information. 


\subsection{Inclination}

{
Combining the inclination value from photometry, $i \approx 40^\circ$, with the 
spectroscopically measured projected rotational velocity $v\sin i$ (Table~\ref{Tab:StarChar}) allows us to determine
in the case of  KIC\,7985370 the star's radius. 
}
Taking the shortest rotational period ($P_{\rm eq}$), 
one arrives at $R = 1.42$--$1.44 R_{\sun}$. 
This (minimal) radius is larger than the ZAMS value of $R\approx 1.1\,R_{\sun}$, but smaller than the 
$R\approx 2\,R_{\sun}$ for a star of $1.5 M_{\sun}$ at 10 Myr. Hence, the photometrically 
derived radius is within the expected range. 
As stated in Sect.~\ref{Results:spots:7765135}, in the case of KIC\,7765135 the photometric 
inclination is badly defined. Therefore, the inclination has been fixed to $i = 75^\circ$, assuming 
the radius to have its ZAMS value.

\subsection{Spot contrast and spot longevity}

Although both stars share the same spectral type and age, there are differences concerning 
the spots. The spots of KIC\,7985370 seem to be darker and longer living than those of KIC\,7765135. 

We would like to remind that a ``spot'' may be in fact a group of smaller 
spots that all together form an active region, which could also {include bright features. } 

For KIC\,7765135, the spot contrast, $\kappa\approx 0.7$, looks rather normal. 
It is similar to the previously studied case of KIC\,8429280 (Paper~I). 
The corresponding temperature contrast, the ratio between spot and photospheric temperature $T_{\rm sp}/T_{\rm ph}$, is 0.9, assuming that the 
``white light'' \kep flux matches the bolometric conditions. The much darker spots, 
$\kappa\approx 0.4$, in the case of KIC\,7985370 defy a simple explanation. 
There is no need for exceptionally small (and therefore dark) spots to prevent spot overlap. 


Apart from a few late F-type stars with low or moderate activity observed by CoRoT \citep{Mosser2009} where spots seem to
be short-lived, there is strong evidence that spots in very active stars like our targets have rather long lives compared to 
the star rotation.  
Active longitudes lasting for months or years have been observed in young stars \citep[e.g., ][]{Collier1995,Hatzes1995,Barnes1998,Huber2009,Lanza2011} 
and in the evolved components of close binary systems, like II~Peg \citep[e.g.][]{Rodono2000,Lindborg2011}.
This does not exclude that individual unresolved spots, which are composing the active region, have shorter evolution 
times, but the photospheric active region seen as an entity endures for a very long time in such cases.

Unlike KIC\,7985370, in the case of KIC\,7765135 mid-latitude spots (30--50$^\circ$) are missing.
This is reminiscent of the spot distribution of two fast-rotating early G dwarfs, He 520 and He 699, of the $\alpha$ Persei cluster studied by
\citet{Barnes1998}. Despite the fact that there is a clear distinction between near-equator and near-pole spots,
with regard to spot lifetimes, no correlation seems to exist between lifetime and latitude,
which is contrary to the case of the rapidly-rotating young AB Dor \citep{Collier1995}, where only
low- and intermediate-latitude spots are long-lived. 

The sudden appearance of a full-grown spot at the beginning of a new quarter of data
is suspicious and hints at an artefact (Fig.~\ref{7985370SpotAreaEvolution}).

\subsection{Differential rotation}
 
Both stars exhibit low-latitude spots as well as high-latitude ones at the time of observation 
making them suitable for studying their latitudinal shear. 
The most robust and important result of the present work is the high degree of surface differential rotation found for both
stars: ${\rm d}\Omega = 0.18$\,rad\,d$^{-1}$. 
{This exceeds threefold the solar value. }

This estimate is rather robust, because any spot model with a few long-lasting spots able to 
reproduce the beating of the light curve must provide a value of equator-to-pole differential rotation 
that exceeds the lower limit of $2\pi/P_{\rm beat}$, irrespective of the number of spots used. 

Inclination has a marginal effect since the periods found in the light curve do not depend on it. 

We remark that the high value ${\rm d}\Omega$ relies on the assumption of spot longevity. 
It is always possible to get an excellent fit with many short-lived spots even for rigid rotation. 



Very different values of differential rotation have been found for \object{HD~171488} 
(\object{V889 Her}), a young ($\sim$50 Myr) Sun.
For this star, which is rotating faster ($P=1.33$\,days) than our targets, a very high solar-type 
differential rotation ${\rm d}\Omega\approx 0.4$--0.5 rad\,d$^{-1}$, with the equator lapping the poles every 12--16 days, 
was found by both \citet{Marsden06} and \citet{Jeffers08}. 
Much weaker values (${\rm d}\Omega\approx 0.04$) were derived instead for the same star by \citet{Jarvi08} and \citet{Kovari11}.
\citet{Huber2009} even claim their data being consistent with no differential rotation. 

\citet{Marsden11} report values of ${\rm d}\Omega$ in the range 0.08--0.45 rad\,d$^{-1}$
for a sample of stars similar to and slightly more massive than the Sun.	
Among these stars, \object{HD~141943}, a 1.3-$M_{\sun}$ star that is still in the PMS phase (age $\sim 17$\,Myr), 
displays values of ${\rm d}\Omega$ ranging from about 0.23 to 0.44 rad\,d$^{-1}$ in different epochs.	
A solar-type differential rotation, ${\rm d}\Omega\approx 0.2$ rad\,d$^{-1}$, was also found by \citet{Waite2011} for 
\object{HD~106506}, a G1\,V-type star ($T_{\rm eff}=5900$\,K) that is very similar to our targets, but it is rotating
faster ($P_{\rm eq} =  1.39$\,days). 
%
Moreover, the Fourier transform technique applied to high-resolution 
spectra of a large sample of F- and early G-type stars indicates that 
differential rotation is rather frequently found \citep{ReinersSchmitt2003,Reiners2006}. 
In their data, there is no clear dependency on the rotation period, but the 
strongest differential rotation, up to $\sim$\,1.0 rad\,d$^{-1}$,  
occurs for periods between 2 and 3 days and values as high as 
$\sim$\,0.7\,rad\,d$^{-1}$ are encountered down to $P\sim 0.5$\,days. 

From ground-based photometry, which is basically devoted to cooler stars, 
a different behaviour, i.e. a differential rotation decreasing with the 
rotation period, seems to emerge \citep[e.g.,][]{Messina03}.
However, the precision of the ground-based light curves does not allow to draw 
firm conclusions and accurate photometry from space, as well as Doppler 
imaging, is needed for settling this point.

For mid-G to M dwarfs, weaker values of the latitudinal shears are generally found. 
In particular, \cite{Barnes05} analyzed with the Doppler imaging technique a small sample of 
active stars in the spectral range G2--M2 finding a trend towards decreasing 
surface differential rotation with decreasing temperature. This suggests that the stellar mass 
must also play a significant role in this respect. 
 The largest values for stars as cool as about 
5000\,K are ${\rm d}\Omega = 0.27$ rad\,d$^{-1}$ found by us in Paper~I 
for KIC\,8429280 (K2\,V, $P=1.16$\,days) and 
${\rm d}\Omega = 0.20$ rad\,d$^{-1}$ found by \citet{Donati03} 
for \object{LQ\,Hya} (K2\,V, $P=1.60$\,days). 
{
The slowly rotating ($P_{\rm eq} = 11.2$\,d) and mildly active K2\,V star \object{$\epsilon$\,Eri} 
exhibits only little differential surface rotation ($0.017\le {\rm d}\Omega\le 0.056$ rad\,d$^{-1}$) 
as a Bayesian reanalysis of the MOST light curve
\citep{Croll2006, Crolletal2006} revealed \citep{Froehlich07}. 
}

{
Thus, there is an indication that a high differential rotation goes along with a high rotation rate. 
}


The high differential rotation that we found for KIC\,7985370 and KIC\,7765135 disagrees with the hydrodynamical model of 
\citet{Kueker11}, which instead predicts a rather low value of ${\rm d}\Omega\approx 0.08$ for an (evolved) solar-mass star 
rotating with a period as short as 1.3\,days. 

{
Surface differential rotation may even vary along the activity cycle. Indeed, certain mean-field dynamo models 
for rapidly rotating cool stars with deep convection zones predict torsional oscillations 
with variations of several percent in differential rotation \citep{Covas2005}. 
Of course, this cannot explain such extreme cases as LQ\,Hya where at times the surface rotation is solid body. 
According to \citet{Lanza2006}, to maintain the strong shear ($\sim\,$0.2 rad\,d$^{-1}$) observed for LQ\,Hya in the year 2000 would 
imply a dissipated power exceeding the star's luminosity. 
}

The differential rotation of rapidly-rotating solar-like stars has been recently investigated on theoretical grounds
by \citet{Hotta2011}. They found that differential rotation approaches the Taylor-Proudman state, i.e. the iso-rotation surfaces
tend to become cylinders parallel to the rotation axis, when stellar rotation is faster than the solar one. In this case, the
differential rotation is concentrated at relatively low latitudes with large stellar angular velocity. They show that
the latitudinal shear (between the equator and latitude $\beta=45\degr$) increases with the angular velocity, in line with our 
results and the recent literature. 

\section{Conclusions}
\label{Sec:Conc}

In this paper, we have studied two Sun-like stars, KIC\,7985370 and KIC\,7765135, by means of high-resolution spectroscopy and 
high-precision \kep photometry.

The high-resolution spectra allowed us to derive, for the first time, their spectral type, astrophysical 
parameters ($T_{\rm eff}, \log g$, [Fe/H]), rotational and heliocentric radial velocities, 
and lithium abundance. All this information, combined with the analysis of the SED and proper motions,
enabled us to infer their distance and kinematics, and to estimate the age of both stars in the 
range 100--200 Myr, although we cannot exclude that they could be as young as 50 Myr.   
Thus, these two sources should be already in the post-T Tauri phase. 

As expected from their young age, both stars were found to be chromospherically active displaying 
filled-in H$\alpha$, H$\beta$, and \ion{Ca}{ii} IRT lines, as well as \ion{He}{i} D$_3$ absorption.
The surface chromospheric fluxes and the X-ray luminosity (for KIC\,7985370), within the ranges found for 
stars with similar $T_{\rm eff}$ and $v\sin i$ in the Pleiades cluster, are just 
below the saturation level \citep{Soderblom1993b}.
The flux ratio of two \ion{Ca}{ii} IRT lines and the Balmer decrement (for KIC\,7765135 only) suggest that
the chromospheric emission is mainly due to optically-thick surface regions analogous to solar plages.

We have applied a robust spot model, based on a Bayesian approach and a MCMC method, to the \kep light
curves which span nearly 229 days and have an unprecedent precision ($\approx 10^{-5}$\,mag).
While seven long-lived spots were needed to perform a reasonable fit (at a 2-mmag level) of the light 
curve of KIC\,7985370, we used up to nine spots in the case of KIC\,7765135 due to a shorter lifetime of its spots. 
Because of the exceptional precision of the \kep photometry it is impossible to reach the 
Bayesian noise floor defined by, e.\,g., the BIC \citep{Schwarz1978} 
without increasing 
significantly the degrees of freedom and, consequently, the non-uniqueness of the solution. 
Provided spots are indeed long-lived, the equator-to-pole value of the shear amounts 
for both stars to $0.18$ rad\,d$^{-1}$.
This is in contrast with the theoretical models of \citet{Kueker11} that predict a moderate 
solar-type differential rotation even for fast-rotating main-sequence stars, 
unless the convection zone is shallower than predicted by the stellar models. 
Our results are instead in line with the scenario proposed by other modelers of a differential rotation that  
increases with the angular velocity \citep{Hotta2011} and that can be also subject to changes along the activity 
cycle \citep{Covas2005,Lanza2006}.

\begin{acknowledgements}
We would like to thank the {\it Kepler} project and the team who created the MAST {\it Kepler} web site and search interfaces.
{We have also to thank an anonymous referee for helpful and constructive comments. }
This work has been partly supported by the Italian {\em Ministero dell'Istruzione, Universit\`a e  Ricerca} (MIUR), which is 
gratefully acknowledged. 
JM\.Z acknowledges the Polish Ministry grant no N\,N203\,405139.
DM and AK acknowledge the Universidad Complutense de Madrid (UCM), 
the Spanish Ministerio de Ciencia e Innovaci\'{o}n (MCINN)
under grants AYA2008-000695 and AYA2011-30147-C03-02, 
and the Comunidad de Madrid under PRICIT project S2009/ESP-1496 (AstroMadrid).
This research made use of SIMBAD and VIZIER databases, operated at the CDS, Strasbourg, France. 
This publication makes use of data products from the Two Micron All Sky Survey, which is a joint 
project of the University of Massachusetts and the Infrared Processing and Analysis Center/California 
Institute of Technology, funded by the National Aeronautics and Space Administration and the National 
Science Foundation.
\end{acknowledgements}

\bibliographystyle{aa}
\bibliography{RefBib}



\end{document}